\documentclass[10pt]{article}

\usepackage[a4paper,top=2.5cm,bottom=2.5cm,left=2.5cm,right=2.5cm]{geometry}
\usepackage{times}
\usepackage{natbib}
\usepackage{amssymb}
\usepackage{amsmath}
\usepackage{amsfonts}
\usepackage{epsfig}
\usepackage{graphicx}
\usepackage[font=small]{caption}

\DeclareMathOperator{\Ima}{Im}

\begin{document}

\title{Stopping of a relativistic electron beam in a plasma irradiated by an
intense laser field}

\author{H.B. NERSISYAN$^{1,2}$\thanks{Address correspondence and reprint requests to: H.B. Nersisyan,
Institute of Radiophysics and Electronics, 0203 Ashtarak, Armenia. E-mail: hrachya@irphe.am}
\textsc{and} C. DEUTSCH$^{3}$ \\ \small{$^{1}$Institute of Radiophysics
and Electronics, Ashtarak, Armenia} \\
\small{$^{2}$Centre of Strong Fields Physics, Yerevan State University, Yerevan, Armenia} \\
\small{$^{3}$LPGP (UMR-CNRS 8578), Universit\'{e} Paris XI, 91405 Orsay, France}}

\date{\today}

\maketitle

\begin{abstract}
The effects of a radiation field (RF) on the interaction process of a relativistic electron beam (REB)
with an electron plasma are investigated. The stopping power of the test electron averaged with a period
of the RF has been calculated assuming an underdense plasma, $\omega_{0} >\omega_{p}$, where $\omega_{0}$
is the frequency of the RF and $\omega_{p}$ is the plasma frequency. In order to highlight the effect of
the radiation field we present a comparison of our analytical and numerical results obtained for nonzero
RF with those for vanishing RF. In particular, it has been shown that the weak RF increases the mean energy
loss for small angles between the velocity of the REB and the direction of polarization of the RF while
decreasing it at large angles. Furthermore, the relative deviation of the energy loss from the field-free
value is strongly reduced with increasing the beam energy. Special case of the parallel orientation of
the polarization of the RF with respect to the beam velocity has been also considered. At high-intensities
of the RF two extreme regimes have been distinguished when the excited harmonics cancel effectively each
other reducing strongly the energy loss or increasing it due to the constructive interference. Moreover,
it has been demonstrated that the energy loss of the ultrarelativistic electron beam increases systematically
with the intensity of the RF exceeding essentially the field-free value.
\end{abstract}

{\textbf{Keywords:} Relativistic electron beam; Radiation field; Stopping power}

\section{Introduction}
\label{sec:1}

The interaction of a charged particles beam with a plasma is an important subject of relevance for many fields
of physics, such as inertial confinement fusion (ICF) driven by ion or electron beams \citep{deu86,deu95,deu01,ava93,cou94,hof08},
high energy density physics \citep{tah05,nel06}, and related astrophysical phenomena \citep{pir05}. This interaction
is also relevant, among others, for the fast ignition scenario (FIS) \citep{tab94,deu96}, where the precompressed
deuterium-tritium (DT) core of a fusion target is to be ignited by a laser-generated relativistic electron beam.
In addition, a promising ICF scheme has been recently proposed \citep{sto96,rot01}, in which the plasma target is
irradiated simultaneously by intense laser and ion beams. Within this scheme several experiments \citep{rot00,ogu00,fra10,fra13,hof10}
have been performed to investigate the interactions of heavy ion and laser beams with plasma or solid targets. An
important aspect of these experiments is the energy loss measurements for the ions in a wide-range of plasma parameters.
It is expected in such experiments that the ion propagation would be essentially affected by the parametric excitation
of the plasma target by means of laser irradiation. This effect has been supported recently by particle-in-cell numerical
simulations \citep{hu11}.

Motivated by the experimental achievements mentioned above, in this paper we present a study of the effects of intense
radiation field (RF) on the interaction of projectile particles with an electron plasma. Previously this has been a subject
of great activity, starting with the pioneering work of Tavdgiridze, Aliev, and Gorbunov \citep{tav70,ali71}. More recently
the need of the comprehensive analysis of the complexity of the beam-matter interaction in the presence of strong laser
fields has stimulated a number of theoretical studies. \citet{ari89} and \citet{ner99} have developed a time-dependent Hamiltonian
formulation to describe the effect of a strong laser field on energy losses of swift ions moving in a degenerate electron gas.
It has been shown that the energy loss is reduced for ions at intermediate velocities, but is increased for slow ions, due to a
resonance process of plasmon excitation in the target assisted by a photon absorption \citep{ari89}. For plasma targets, on
the other hand, it has been obtained that the energy loss is reduced in the presence of a laser field when the projectile
velocity is smaller than the electron thermal velocity \citep{ari90,ner11}. Moreover, the reduction of the energy loss rate
increases with decreasing the angle between the projectile velocity and the polarization vector of the laser field
\citep{ner11}. Next, an interesting effect has been obtained by \citet{ako97} considering the oscillatory motion of the
light projectile particles in the laser field. In this case the energy loss depends on the projectile charge sign and mass
and, therefore, results in a different stopping powers for electrons and positrons. In recent years special attention has been
also paid to the regimes of high intensities of the laser field when the quiver velocity of the plasma electron exceeds the
projectile velocity \citep{ner99,ner11}. It has been shown that the projectile particles can be even accelerated in a plasma
due to the parametric effects excited by an intense laser field. Furthermore, \citet{wan02,wan12} have studied
the influence of a high-intensity laser field on the Coulomb explosion and stopping power for a swift H$^{+}_{3}$ cluster ion
in a plasma target. They have demonstrated that the laser field affects the correlation between the ions and contributes to
weaken the wake effect as compared to the laser-free case.

All papers mentioned above deal with a nonrelativistic projectile particles interacting with a laser irradiated plasma. In the
present paper we study more general configurations thus considering the interaction of relativistic particles with an electron plasma.
Our objective is then to study some aspects of the energy loss of relativistic particles which have not been considered previously.
As an important motivation of the present paper is the research on the topic of the FIS for inertial confinement fusion
\citep{deu96,tab94} which involves the interaction of a laser-generated relativistic electron beam (REB) with a plasma. Although
the concept of the FIS implies an overdense plasma and the propagation of a relativistic electron beam from the border of a
precompressed target to the dense core occurs without crossing the laser beam, the target plasma is assumed to be parametrically
excited by the RF through high-harmonics generation. In this situation, the presence of the RF as well as the relativistic effects
of the beam can dramatically change the main features of the standard stopping capabilities of the plasma.

The plan of the paper is as follows. In Section~\ref{sec:2}, we outline the theoretical formalism for the electromagnetic response
of a plasma due to the motion of a relativistic charged particles beam in a plasma in the presence of an intense RF. The stopping
of the relativistic beam is formulated in Section~\ref{sec:3}. In this section we calculate the effects of the RF on the mean energy
loss (stopping power) of the test particle considering two somewhat distinct cases with a weak (Sec.~\ref{sec:3-1}) and intense
(Sec.~\ref{sec:3-2}) radiation field as well as a wide range for the test particle energies, extending from nonrelativistic up to the
ultrarelativistic energies. The dynamics of the electron plasma is treated by a system of fluid equations which results in a high-frequency
approximation for the dielectric response function. The results are summarized in Section~\ref{sec:4}.

\section{Solution of the basic equations}
\label{sec:2}

The whole interaction process of the relativistic charged particles beam with plasma involves the
energy loss and the charge states of the projectile particle and -- as an additional aspect -- the
ionization and recombination of the projectile driven by the RF and the collisions with the plasma
particles. A complete description of the interaction of the beam requires a simultaneous treatment
of all these effects including, in particular, in the case of the structured particles the effect
of the charge equilibration on the energy loss process. In this paper, we do not discuss the charge
state evolution of the projectiles under study, but concentrate on the RF effects on the energy loss
process assuming an equilibrium charge state of the projectile particle with an effective charge $Ze$.
This is motivated by the fact that the charge equilibration occurs in time scales, which are usually
much smaller than the time of passage of the beam through target. In addition, the effects mentioned
above are not important in the case of the relativistic electron beams.

The problem is formulated using the hydrodynamic model of the plasma and includes the effects of the
RF in a self-consistent way. The external RF is treated in the long-wavelength limit, and the plasma
electrons are considered nonrelativistic. These are good approximations provided that (i) the wavelength
of the RF ($\lambda _{0}=2\pi c/\omega _{0}$) is much larger than the Debye screening length $\lambda_{D}
=v_{\mathrm{th}}/\omega _{p}$ (with $v_{\mathrm{th}}$ the thermal velocity of the electrons and $\omega
_{p} $ the plasma frequency), and (ii) the \textquotedblleft quiver velocity\textquotedblright\ of the
electrons in the RF $v_{E}=eE_{0}/m\omega_{0}$ (where $\mathbf{E}_{0}$ is the amplitude of the RF) is
much smaller than the speed of light $c$. These conditions can be alternatively written as (i) $\omega_{0}
/\omega _{p}\ll 2\pi c/v_{\mathrm{th}}$, and (ii) $W_{L}\ll \frac{1}{2}n_{e}mc^{3}(\omega _{0}/\omega_{p})^{2}$,
where $W_{L}=cE_{0}^{2}/8\pi$ is the intensity of the RF. As an estimate in the case of a dense plasma,
with electron density $n_{e}=10^{22}$~cm$^{-3}$, we get $\frac{1}{2}n_{e}mc^{3}\simeq 1.2\times 10^{19}$~W/cm$^{2}$.
Thus the conditions (i) and (ii) are well above the values obtained with currently available high-power
sources of the RF, and these approximations are well justified.

The response of the plasma to the incoming relativistic beam is governed by the hydrodynamic equations
for the density $n(\mathbf{r},t)$ and the velocity $\mathbf{v}(\mathbf{r},t)$ of the plasma electrons as
well as by the Maxwell equations for the electromagnetic fields. Thus,
\begin{eqnarray}
&&\frac{\partial n}{\partial t}+\nabla (n\cdot \mathbf{v}) =0 ,  \label{eq:1} \\
&&\frac{\partial \mathbf{v}}{\partial t}+(\mathbf{v}\cdot \nabla )\mathbf{v}
=-\frac{e}{m}\left[ \mathbf{E}_{0}(t)+\mathbf{E}+\frac{1}{c}\left[ \mathbf{%
v}\times \mathbf{B}\right] \right] ,  \label{eq:2}
\end{eqnarray}%
where $\mathbf{E}_{0}(t)=\mathbf{E}_{0}\sin (\omega _{0}t)$ is the RF, and $\mathbf{E}$ and $\mathbf{B}$
are the self-consistent electromagnetic fields which are determined by the Maxwell equations,
\begin{eqnarray}
&&\nabla \times \mathbf{E} =-\frac{1}{c}\frac{\partial \mathbf{B}}{\partial t%
},\qquad \nabla \times \mathbf{B}=\frac{4\pi }{c}\left( \mathbf{j}_{0}+%
\mathbf{j}\right) +\frac{1}{c}\frac{\partial \mathbf{E}}{\partial t} , \label{eq:3} \\
&&\nabla \cdot \mathbf{E} =-4\pi e\left( n-n_{e}\right) +4\pi \rho
_{0},\qquad \nabla \cdot \mathbf{B}=0 .  \label{eq:4}
\end{eqnarray}%
Here $\mathbf{j}=-en\mathbf{v}$ is the induced current density, $\rho_{0}$ and $\mathbf{j}_{0}$ are the
charge and current densities for the beam, respectively, $n_{e}$ is the equilibrium plasma density in an
unperturbed state. As mentioned above we consider a relativistic beam of charged particles and, therefore,
the influence of the RF $\mathbf{E}_{0}(t)$ on the beam is ignored.

Consider now the solution of Eqs.~\eqref{eq:1}-\eqref{eq:4} for the beam-plasma system in the presence of
the high-frequency RF. In an unperturbed state (i.e., neglecting the self-consistent electromagnetic fields
$\mathbf{E}$ and $\mathbf{B}$ in Eq.~\eqref{eq:2} and assuming the homogeneous initial state) the plasma
velocity satisfies the equation $\dot{\mathbf{u}}_{e}(t)=-\frac{e}{m}\mathbf{E}_{0}(t)$ which yields the
equilibrium velocity for the plasma electrons, $\mathbf{u}_{e}(t)=\mathbf{v}_{E}\cos (\omega_{0}t)$. Here
$\mathbf{v}_{E}=e\mathbf{E}_{0}/m\omega _{0}$ and $\mathbf{a}=e\mathbf{E}_{0}/m\omega_{0}^{2}$ are the quiver
velocity and the oscillation amplitude of the plasma electrons, respectively, driven by the RF.

Next we consider the linearized hydrodynamic and Maxwell's equations for the plasma and electromagnetic fields,
respectively. For sufficiently small perturbations, we assume $\mathbf{v}(\mathbf{r},t)=\mathbf{u}_{e}(t)+\delta
\mathbf{v}(\mathbf{r},t)$, $n(\mathbf{r},t)=n_{e}+\delta n(\mathbf{r},t)$, (with $\delta n\ll n_{e}$ and
$\delta v\ll u_{e}$). Thus, introducing the Fourier transforms $\delta n(\mathbf{k},t)$, $\delta \mathbf{v}
(\mathbf{k},t)$, $\mathbf{E}(\mathbf{k},t)$, and $\mathbf{B}(\mathbf{k},t)$ with respect to the coordinate
$\mathbf{r}$, the linearized hydrodynamic equations read
\begin{eqnarray}
&&\left[ \frac{\partial }{\partial t}+i\left( \mathbf{k}\cdot \mathbf{u}%
_{e}(t)\right) \right] \delta n(\mathbf{k},t) =-in_{e}\left( \mathbf{k}%
\cdot \delta \mathbf{v}(\mathbf{k},t)\right) ,  \label{eq:5} \\
&&\left[ \frac{\partial }{\partial t}+i\left( \mathbf{k}\cdot \mathbf{u}%
_{e}(t)\right) \right] \delta \mathbf{v}(\mathbf{k},t) =-\frac{e}{m}\left[
\mathbf{E}(\mathbf{k},t)+\frac{1}{c}\left[ \mathbf{u}_{e}(t)\times \mathbf{B}%
(\mathbf{k},t)\right] \right] .  \label{eq:6}
\end{eqnarray}%
In order to solve Eqs.~\eqref{eq:5} and \eqref{eq:6} it is convenient to introduce instead of the Fourier
transform $\mathbf{P}(\mathbf{k},t)$ a new unknown function $\widetilde{\mathbf{P}}(\mathbf{k},t)$ via the relation
\citep{ner12}
\begin{equation}
\widetilde{\mathbf{P}}(\mathbf{k},t)=\mathbf{P}(\mathbf{k},t)e^{i\zeta \sin
(\omega _{0}t)},  \label{eq:7}
\end{equation}%
where $\zeta =\mathbf{k}\cdot \mathbf{a}$, $\mathbf{P}(\mathbf{k},t)$ represents one of the quantities $\delta
n(\mathbf{k},t)$, $\delta \mathbf{v}(\mathbf{k},t)$, $\mathbf{E}(\mathbf{k},t)$, and $\mathbf{B}(\mathbf{k},t)$.
Substituting this relation for the quantities $\delta n(\mathbf{k},t)$, $\delta \mathbf{v}(\mathbf{k},t)$,
$\mathbf{E}(\mathbf{k},t)$, and $\mathbf{B}(\mathbf{k},t)$ into Eqs.~\eqref{eq:5} and \eqref{eq:6} it is easy to
see that the obtained equation for the unknown function $\widetilde{\mathbf{P}}(\mathbf{k},t)$ constitutes an
equation with periodic coefficients \citep{ner12}. Therefore, we introduce the decompositions
\begin{eqnarray}
&&\mathbf{P}(\mathbf{k},t) =\int_{-\infty }^{\infty }d\omega e^{-i\omega
t}\sum_{\ell =-\infty }^{\infty }\mathbf{P}_{\ell }(\mathbf{k},\omega
)e^{-i\ell \omega _{0}t} ,  \label{eq:8} \\
&&\widetilde{\mathbf{P}}(\mathbf{k},t) =\int_{-\infty }^{\infty }d\omega
e^{-i\omega t}\sum_{\ell =-\infty }^{\infty }\widetilde{\mathbf{P}}_{\ell }(%
\mathbf{k},\omega )e^{-i\ell \omega _{0}t} ,  \label{eq:9}
\end{eqnarray}%
where $\mathbf{P}_{\ell }(\mathbf{k},\omega )$ and $\widetilde{\mathbf{P}}_{\ell }(\mathbf{k},\omega )$ are the
corresponding amplitudes of the $\ell$th harmonic. It is also useful to find the connection between these
amplitudes. This can be done using the Fourier series representation of the exponential function $e^{i\zeta\sin
(\omega_{0}t)}$ as well as the summation formula $\sum_{r=-\infty }^{\infty }J_{r}(\zeta )J_{r+\ell }(\zeta)=
\delta_{\ell 0}$ for the Bessel functions \citep{gra80} which yields
\begin{eqnarray}
&&\widetilde{\mathbf{P}}_{\ell }(\mathbf{k},\omega ) = \sum_{s=-\infty
}^{\infty }J_{s-\ell }(\zeta )\mathbf{P}_{s}(\mathbf{k},\omega ) , \label{eq:10} \\
&&\mathbf{P}_{\ell }(\mathbf{k},\omega ) = \sum_{s=-\infty }^{\infty }J_{\ell
-s}(\zeta )\widetilde{\mathbf{P}}_{s}(\mathbf{k},\omega ) .  \label{eq:11}
\end{eqnarray}%
Here $J_{\ell }$ is the Bessel function of the first kind and of the $\ell $th order. Using these results from
Eqs.~\eqref{eq:5} and \eqref{eq:6} it is straightforward to obtain
\begin{eqnarray}
&&\delta \widetilde{n}_{\ell }(\mathbf{k},\omega ) = \frac{n_{e}}{\Omega_{\ell}}
\left( \mathbf{k}\cdot \delta \widetilde{\mathbf{v}}_{\ell
}(\mathbf{k},\omega )\right) ,  \label{eq:12} \\
&&\delta \widetilde{\mathbf{v}}_{\ell }(\mathbf{k},\omega ) = -\frac{ie}{%
m\Omega_{\ell}}\widetilde{\mathbf{E}}_{\ell }(\mathbf{k},\omega
),  \label{eq:13}
\end{eqnarray}%
where $\Omega _{\ell}=\omega +\ell\omega _{0}$ and we have neglected the term of the order of $v_{E}/c$ (see
the last term in Eq.~\eqref{eq:6}) according to the approximation (ii). Using the transformation formulas
\eqref{eq:10} and \eqref{eq:11} from the system of equations \eqref{eq:12} and \eqref{eq:13} it is easy to obtain
\begin{eqnarray}
&&\delta n_{\ell }(\mathbf{k},\omega ) = -\frac{in_{e}e}{m}\sum_{s,r=-\infty
}^{\infty }\frac{1}{\Omega _{r}^{2}}J_{\ell -r}(\zeta )J_{s-r}(\zeta )\left(
\mathbf{k}\cdot \mathbf{E}_{s}(\mathbf{k},\omega )\right) ,  \label{eq:14} \\
&&\delta \mathbf{v}_{\ell }(\mathbf{k},\omega ) = -\frac{ie}{m}%
\sum_{s,r=-\infty }^{\infty }\frac{1}{\Omega _{r}}J_{\ell -r}(\zeta
)J_{s-r}(\zeta )\mathbf{E}_{s}(\mathbf{k},\omega ) .  \label{eq:15}
\end{eqnarray}%

Next let us evaluate the induced current density which is given by $\delta\mathbf{j}=-e[n_{e}\delta \mathbf{v}+
\mathbf{u}_{e}(t)\delta n]$ in the coordinate space. Employing Eqs.~\eqref{eq:8}, \eqref{eq:9}, \eqref{eq:14} and
\eqref{eq:15} for the Fourier transform of the induced current density we obtain
\begin{eqnarray}
&&\delta \mathbf{j}_{\ell }(\mathbf{k},\omega ) = -e\left\{ n_{e}\delta
\mathbf{v}_{\ell }(\mathbf{k},\omega )+\frac{\mathbf{v}_{E}}{2}\left[ \delta
n_{\ell +1}(\mathbf{k},\omega )+\delta n_{\ell -1}(\mathbf{k},\omega )\right] \right\}  \label{eq:16} \\
&&=\frac{i\omega _{p}^{2}}{4\pi }\sum_{s,r=-\infty }^{\infty }\frac{1}{%
\Omega _{r}}J_{\ell -r}(\zeta )J_{s-r}(\zeta )\left[ \mathbf{E}_{s}(\mathbf{k%
},\omega )+\frac{(\ell -r)\omega _{0}}{\Omega _{r}}\frac{\mathbf{a}}{\zeta }%
\left( \mathbf{k}\cdot \mathbf{E}_{s}(\mathbf{k},\omega )\right) \right] ,
\nonumber
\end{eqnarray}%
where $\omega _{p}^{2}=4\pi n_{e}e^{2}/m$ is the plasma frequency.

From the Maxwell equations we express the magnetic field through the electric field. In terms of the amplitudes
of the $\ell$th harmonics [see definition given by Eqs.~\eqref{eq:8} and \eqref{eq:9}] the electric field is
determined by the system of equations
\begin{eqnarray}
&&\left( k^{2}-\frac{\Omega _{\ell }^{2}}{c^{2}}\right) \mathbf{E}_{\ell }(%
\mathbf{k},\omega )-\mathbf{k}\left( \mathbf{k}\cdot \mathbf{E}_{\ell }(%
\mathbf{k},\omega )\right) =\frac{4\pi i\Omega _{\ell }}{c^{2}}\left[ \delta
_{\ell 0}\mathbf{j}_{0}(\mathbf{k},\omega )+\delta \mathbf{j}_{\ell }(%
\mathbf{k},\omega )\right] ,  \label{eq:17} \\
&&\left( \mathbf{k}\cdot \mathbf{E}_{\ell }(\mathbf{k},\omega )\right) =4\pi
i\left[ e\delta n_{\ell }(\mathbf{k},\omega )-\delta _{\ell 0}\rho _{0}(%
\mathbf{k},\omega )\right] .  \label{eq:18}
\end{eqnarray}%
Here $\mathbf{j}_{0}(\mathbf{k},\omega )$ and $\rho _{0}(\mathbf{k},\omega )$ are the ordinary Fourier transforms
of the current and charge densities of the beam, respectively. Note that the $\ell $th harmonics of these quantities
are given by $\delta_{\ell 0}\mathbf{j}_{0}(\mathbf{k},\omega )$ and $\delta_{\ell 0}\rho_{0}(\mathbf{k},\omega )$.
Also the longitudinal part of the electric field is determined by the Poisson equation \eqref{eq:18}.

Insertion of Eq.~\eqref{eq:16} into Eq.~\eqref{eq:17} results in
\begin{eqnarray}
&&\left[ k^{2}-\frac{\Omega _{\ell }^{2}}{c^{2}}\varepsilon (\Omega _{\ell })%
\right] \mathbf{E}_{\ell }(\mathbf{k},\omega )-\mathbf{k} (\mathbf{k}%
\cdot \mathbf{E}_{\ell }(\mathbf{k},\omega ))  \nonumber \\
&&=-\frac{\omega _{p}^{2}}{c^{2}}\sum_{s,r=-\infty }^{\infty }\frac{(\ell
-r)\omega _{0}}{\Omega _{r}}J_{\ell -r}(\zeta )J_{s-r}(\zeta )\left[ \mathbf{%
E}_{s}(\mathbf{k},\omega )+\frac{\Omega _{\ell }}{\Omega _{r}}\frac{\mathbf{a%
}}{\zeta }\left( \mathbf{k}\cdot \mathbf{E}_{s}(\mathbf{k},\omega )\right) \right]  \label{eq:19} \\
&&+\frac{4\pi i\Omega _{\ell }}{c^{2}}\delta _{\ell 0}\mathbf{j}_{0}(\mathbf{k},\omega ) ,  \nonumber
\end{eqnarray}%
where $\varepsilon (\omega )=1-\omega _{p}^{2}/\omega ^{2}$ is the longitudinal dielectric function of a plasma.
The obtained expression \eqref{eq:19} completely determines the electromagnetic response in the beam-plasma system
in the presence of the RF. The resulting equation represents a coupled and infinite system of linear equations for
the quantities $\mathbf{E}_{\ell }(\mathbf{k},\omega )$ (with $\ell =0,\pm 1,\pm 2,\ldots$). The (infinite) determinant
of this system determines the dispersion equation for the electromagnetic plasma modes in the presence of the RF.
It should be also emphasized that the hydrodynamic description of a plasma is justified at high frequencies when
$|\omega+ \ell\omega_{0}| \gg kv_{\mathrm{th}}$. In particular, this condition requires that the velocity of the
projectile particle should exceed the thermal velocity $v_{\mathrm{th}}$ of the plasma electrons.

Equation~\eqref{eq:19} can be further simplified excluding there the longitudinal part $(\mathbf{k}\cdot \mathbf{E}_{\ell})$
of the electric field by means of the Poisson equation \eqref{eq:18} together with the induced charge density,
Eq.~\eqref{eq:14}. In the first step we multiply both sides of Eq.~\eqref{eq:18} by $J_{\ell -q}(\zeta )J_{m-q}(\zeta )$
and sum up over the harmonic numbers $\ell $ and $q$. Using the summation formula involving two Bessel functions (see above
Eq.~\eqref{eq:10}) this results in
\begin{equation}
\sum_{s,r=-\infty }^{\infty }\varepsilon (\Omega _{r})J_{\ell -r}(\zeta
)J_{s-r}(\zeta )\left( \mathbf{k}\cdot \mathbf{E}_{s}(\mathbf{k},\omega
)\right) =-4\pi i\delta _{\ell 0}\rho _{0}(\mathbf{k},\omega ).
\label{eq:20}
\end{equation}
Similarly repeating the first step for the latter expression for the longitudinal electric field one finally obtains
\begin{equation}
\left( \mathbf{k}\cdot \mathbf{E}_{\ell }(\mathbf{k},\omega )\right) =-4\pi
i\rho _{0}(\mathbf{k},\omega )\sum_{r=-\infty }^{\infty }\frac{(-1)^{r}}{%
\varepsilon (\Omega _{r})}J_{\ell -r}(\zeta )J_{r}(\zeta ) .
\label{eq:21}
\end{equation}
This expression has been previously obtained and studied by many authors \citep{tav70,ali71,ari89,ari90,ako97,ner99,ner11}
assuming nonrelativistic beam of charged particles. Thus, inserting Eq.~\eqref{eq:21} into Eq.~\eqref{eq:19} we finally
obtain
\begin{eqnarray}
&&\left[ k^{2}-\frac{\Omega _{\ell }^{2}}{c^{2}}\varepsilon (\Omega _{\ell })%
\right] \mathbf{E}_{\ell }(\mathbf{k},\omega )=\frac{4\pi i\omega }{c^{2}}%
\delta _{\ell 0}\mathbf{j}_{0}\left( \mathbf{k},\omega \right)  \nonumber \\
&&-\frac{\omega _{p}^{2}}{c^{2}}\sum_{s,r=-\infty }^{\infty }\frac{(\ell
-r)\omega _{0}}{\Omega _{r}}J_{\ell -r}(\zeta )J_{s-r}(\zeta )\mathbf{E}_{s}(%
\mathbf{k},\omega )  \label{eq:22} \\
&&-4\pi i\rho _{0}(\mathbf{k},\omega )\sum_{r=-\infty }^{\infty }(-1)^{r}%
\frac{\boldsymbol{\chi }_{\ell r}(\mathbf{k},\omega )}{\varepsilon (\Omega
_{r})}J_{\ell -r}(\zeta )J_{r}(\zeta ),  \nonumber
\end{eqnarray}%
where we have introduced the vector quantity
\begin{equation}
\boldsymbol{\chi }_{\ell r}(\mathbf{k},\omega )=\mathbf{k}-(\ell -r)\omega
_{0}\Omega _{\ell }\left[ 1-\varepsilon (\Omega _{r})\right] \frac{\mathbf{a}%
}{c^{2}\zeta } .
\label{eq:23}
\end{equation}

As mentioned above Eq.~\eqref{eq:19} as well as its equivalent Eq.~\eqref{eq:22} represent an infinite system of the
coupled linear equations for the amplitudes $\mathbf{E}_{\ell}(\mathbf{k},\omega )$ which, in general, do not allow
an analytic solution. Next, we will make an \emph{ad hoc} assumption neglecting the second term in the right-hand side
of Eq.~\eqref{eq:22} which contains an infinite sum over the harmonics $\mathbf{E}_{s}$. In this case the solution of
Eq.~\eqref{eq:22} is trivial and it is given by
\begin{eqnarray}
&&\mathbf{E}_{\ell }(\mathbf{k},\omega ) = \frac{4\pi i}{k^{2}-(\Omega_{\ell }^{2}/c^{2})
\varepsilon (\Omega_{\ell })}\bigg[ \frac{\omega }{c^{2}%
}\delta _{\ell 0}\mathbf{j}_{0}(\mathbf{k},\omega )  \label{eq:24} \\
&& -\rho _{0}(\mathbf{k},\omega )\sum_{r=-\infty }^{\infty }(-1)^{r}%
\frac{\boldsymbol{\chi }_{\ell r}(\mathbf{k},\omega )}{\varepsilon (\Omega
_{r})}J_{\ell -r}(\zeta )J_{r}(\zeta )\bigg] .  \nonumber
\end{eqnarray}%
Now let us examine the physical consequences of the neglect of the second term in Eq.~\eqref{eq:22}. For this purpose
we insert the last expression into the second term of the right-hand side of Eq.~\eqref{eq:22}. Then it is straightforward
to see that this term corresponds to the \v{C}erenkov radiation (or absorbtion) in a plasma by a moving charged particles
beam. Note that this effect is absent in the field-free case since the physical conditions of the \v{C}erenkov radiation
cannot be fulfilled in this case. Thus the RF essentially changes the dispersion properties of the plasma and the
\v{C}erenkov effect becomes now possible. In particular, an inspection of a general expression \eqref{eq:22} shows that
in a long wavelength limit the emission of the transverse electromagnetic waves occurs at the frequencies $\omega_{\ell }
=\ell\omega_{0}\beta /(1-\beta )$ for $\ell \geqslant 1$ (emission), and at $\omega =|\ell |\omega _{0}\beta /(1+\beta )$
for $\ell \leqslant -1$ (absorbtion), where $\beta =u/c$ and $u$ are the relativistic factor and the velocity of the beam,
respectively. Moreover, the ratio of the neglected term, related to the \v{C}erenkov effect, to the contribution given by
Eq.~\eqref{eq:24} is of the order of $(\omega_{p}/\omega _{0})(v_{E}/u)^{2}<1$ and the neglect of the second term in
Eq.~\eqref{eq:22} is justified at the high-frequencies of the RF ($\omega_{0}\gg \omega _{p}$). In the next sections the
further calculations will be done using the approximate expression \eqref{eq:24}.

Before closing this section let us consider briefly some limiting cases of Eq.~\eqref{eq:24}. For instance, in the
electrostatic limit (assuming that $c\to \infty $) from Eq.~\eqref{eq:24} one obtains
\begin{equation}
\mathbf{E}_{\ell }(\mathbf{k},\omega )=-\frac{4\pi i\mathbf{k}}{k^{2}}\rho
_{0}(\mathbf{k},\omega )\sum_{r=-\infty }^{\infty }\frac{(-1)^{r}}{%
\varepsilon (\Omega _{r})}J_{\ell -r}(\zeta )J_{r}(\zeta )
\label{eq:25}
\end{equation}%
which confirms Eq.~\eqref{eq:21}. In the limit of the vanishing radiation field with $a\to 0$, Eq.~\eqref{eq:24} yields
$\mathbf{E}_{\ell }(\mathbf{k},\omega )=\mathbf{E}(\mathbf{k},\omega )\delta_{\ell 0}$, where
\begin{equation}
\mathbf{E}(\mathbf{k},\omega )=\frac{4\pi i}{k^{2}-(\omega^{2}/c^{2})%
\varepsilon (\omega )}\left[ \frac{\omega }{c^{2}}\mathbf{j}_{0}(
\mathbf{k},\omega ) -\frac{\mathbf{k}}{\varepsilon (\omega )}\rho _{0}(%
\mathbf{k},\omega )\right] .
\label{eq:26}
\end{equation}%
It is seen that in the last expression $\mathbf{E}(\mathbf{k},\omega )$ is the Fourier transform of the electric
field created in a medium by an external charge with the density $\rho_{0}(\mathbf{k},\omega )$ and current
$\mathbf{j}_{0}(\mathbf{k},\omega )$ \citep{ale84}.

\section{Stopping power}
\label{sec:3}

With the basic results presented in the previous Sec.~\ref{sec:2}, we now take up the main topic of this paper,
namely the investigation of the stopping power of a relativistic charged particle beam in a plasma irradiated by an intense
laser field. For further progress the charge and current densities of the beam in Eq.~\eqref{eq:24} should be
specified. Furthermore, we will assume that these quantities are given functions of the coordinates and time. In
particular, for a beam of charged particles having a total charge $Ze$ and moving with a constant velocity
$\mathbf{u}$, the charge and current densities are determined by $\rho_{0}(\mathbf{r},t)=ZeQ(\mathbf{r}-\mathbf{u}t)$
and $\mathbf{j}_{0}(\mathbf{r},t)=\mathbf{u}\rho _{0}(\mathbf{r},t)$, respectively. Here $Q(\mathbf{r})$ is the
spatial distribution of the charge in the beam with the normalization $\int Q(\mathbf{r})d\mathbf{r}=1$. Accordingly,
the Fourier transforms of these quantities are $\rho _{0}(\mathbf{k},\omega )=[Ze/(2\pi )^{3}]Q(\mathbf{k})\delta
(\omega -\mathbf{k}\cdot \mathbf{u})$, $\mathbf{j}_{0}(\mathbf{k},\omega )=\mathbf{u}\rho _{0}(\mathbf{k},\omega )$,
and $Q(\mathbf{k})$ is the Fourier transform of the distribution $Q(\mathbf{r})$ (form-factor of the beam). Note
that $Q(\mathbf{k})$ is normalized in a such way that $Q(\mathbf{k}=0)=1$. In particular, for the Gaussian beam
with a charge distribution $Q(\mathbf{r}) =(\pi^{3/2}L_{\parallel} L^{2}_{\perp})^{-1}\exp(-z^{2}/L^{2}_{\parallel})
\exp(-r^{2}_{\perp}/L^{2}_{\perp})$, we obtain $Q(\mathbf{k}) =\exp (-k^{2}_{\parallel}L^{2}_{\parallel}/4)
\exp (-k^{2}_{\perp}L^{2}_{\perp}/4)$. Here $L_{\parallel}$ and $L_{\perp}$ are the characteristic sizes of the
beam along and transverse to the direction of motion, respectively.

From Eqs.~\eqref{eq:8} and \eqref{eq:24} it is straightforward to calculate the time average (with respect to the
period $2\pi /\omega_{0}$ of the laser field) of the stopping field acting on the beam. Then, the averaged stopping
power (SP) of the beam becomes
\begin{eqnarray}
&&S \equiv -\frac{1}{u}\int \left\langle \rho _{0}(\mathbf{r},t)\left(
\mathbf{u}\cdot \mathbf{E}(\mathbf{r},t)\right) \right\rangle d\mathbf{r} \nonumber \\
&& =\frac{2Z^{2}e^{2}}{(2\pi )^{2}u}\Ima\int |Q(\mathbf{k})|^{2}d\mathbf{k}%
\int_{-\infty }^{\infty }\delta (\omega -\mathbf{k}\cdot \mathbf{u})\frac{%
\omega d\omega }{D(k,\omega )}  \label{eq:27} \\
&& \times \left\{ \beta ^{2}-\sum_{\ell =-\infty }^{\infty }\frac{J_{\ell
}^{2}(\zeta )}{\varepsilon (\Omega _{\ell })}\left[ 1+\ell \omega _{0}\left[
1-\varepsilon (\Omega _{\ell })\right] \frac{(\mathbf{u}\cdot \mathbf{a})}{%
c^{2}\zeta }\right] \right\} ,  \nonumber
\end{eqnarray}%
where $D(k,\omega )=k^{2}- (\omega ^{2}/c^{2})\varepsilon (\omega )$ and $\beta =u/c$ is the relativistic factor of
the beam. In Eq.~\eqref{eq:27} the symbol $\langle \ldots\rangle $ denotes an average with respect to the period of
the laser field. Hence, the SP depends on the beam velocity $\mathbf{u}$, the frequency $\omega _{0}$ and the intensity
$W_{L}=cE_{0}^{2}/8\pi $ of the RF (the intensity dependence is given through the quiver amplitude $\mathbf{a}$).
Moreover, since the vector $\mathbf{k}$ in Eq.~\eqref{eq:27} is integrated over the angular variables, $S$ becomes
also a function of the angle $\vartheta$ between the velocity $\mathbf{u}$, and the direction of RF polarization,
represented by $\mathbf{a}$. In addition, it is well known that within classical description an upper cut-off parameter
$k_{\max }=1/r_{\min}$ (where $r_{\min}$ is the effective minimum impact parameter or the distance of closest approach)
must be introduced in Eq.~\eqref{eq:27}
to avoid the logarithmic divergence at large $k$. This divergence corresponds to the incapability of the classical
perturbation theory to treat close encounters between the projectile particles and the plasma electrons properly. For
$r_{\min}$, we use the effective minimum impact parameter excluding hard Coulomb collisions with a scattering angle
larger than $\pi /2$. The resulting cut-off parameter $k_{\max}\simeq m(u^{2}+v_{\mathrm{th}}^{2})/|Z|e^{2}$ is well
known for energy loss calculations (see, e.g., \citet{zwi99}; \citet{ner07} and references therein). Here $v_{\mathrm{th}}$
is the thermal velocity of the electrons. In particular, at high-velocities this cut-off parameter reads as $k_{\max}\simeq
mu^{2}/|Z|e^{2}$. When, however, $u>2|Z|e^{2}/\hbar$, the de Broglie wavelength exceeds the classical distance of closest
approach. Under these circumstances we choose $k_{\max}=2mu/\hbar$.

Equation~\eqref{eq:27} contains two types of the contributions. The first one is proportional to the imaginary part
of the inverse of the longitudinal dielectric function $\varepsilon (\Omega_{\ell})$ and is the energy loss of the
beam due to the excitations of the longitudinal plasma modes. The second contribution comes from the imaginary part
of $[D(k,\omega )]^{-1}$, where $D(k,\omega )$ has been introduced above and determines the energy loss of the beam
due to the excitations of the transverse plasma modes. Therefore, due to the resonant nature of the integral in
Eq.~\eqref{eq:27}, we may replace the imaginary parts of the functions $[\varepsilon (\Omega_{\ell })]^{-1}$ and
$[D(k,\omega )]^{-1}$ using the property of the Dirac $\delta$ function. However, as discussed in the preceding section,
the terms containing the factor $\delta (D(k,\omega ))$ with zero harmonic number ($\ell =0$) correspond to the usual
\v{C}erenkov effect which is absent here. Neglecting such terms we finally arrive at
\begin{eqnarray}
&&S = \frac{Z^{2}e^{2}}{2\pi u}\int |Q(\mathbf{k})|^{2}d\mathbf{k}%
\int_{-\infty }^{\infty }\delta (\omega -\mathbf{k}\cdot \mathbf{u})\frac{%
\omega d\omega }{D(k,\omega )}  \label{eq:28} \\
&&\times \sum_{\ell =-\infty }^{\infty }\frac{|\Omega _{\ell }|}{\Omega
_{\ell }}J_{\ell }^{2}(\zeta )\left[ 1+\frac{\omega _{p}^{2}}{\Omega _{\ell
}^{2}}\frac{(\mathbf{u}\cdot \mathbf{a})}{c^{2}\zeta }\ell \omega _{0}\right]
\delta \left( \varepsilon (\Omega _{\ell })\right) .  \nonumber
\end{eqnarray}%
By comparison, the SP in the absence of the RF is given by \citep{ner99,ner11}
\begin{equation}
S_{0}=\frac{Z^{2}e^{2}}{2\pi u}\int |Q(\mathbf{k})|^{2}\frac{d\mathbf{k}}{%
k^{2}}\int_{-\infty }^{\infty }\delta (\omega -\mathbf{k}\cdot \mathbf{u}%
)\delta \left( \varepsilon (\omega )\right) |\omega |d\omega
\label{eq:29}
\end{equation}%
which after straightforward integration for the point-like particles (with $Q(\mathbf{k})=1$) yields
\begin{equation}
S_{0}=\frac{Z^{2}e^{2}\omega _{p}^{2}}{2u^{2}}\ln \left( 1+\frac{k_{\max
}^{2}u^{2}}{\omega _{p}^{2}}\right) \simeq \frac{Z^{2}e^{2}\omega _{p}^{2}}{%
u^{2}}\ln \frac{k_{\max }u}{\omega _{p}} .
\label{eq:30}
\end{equation}
In the presence of the RF, the SP $S_{0}$ is modified and is given by the $\ell =0$ term in Eq.~\eqref{eq:28}
(\textquotedblleft no photon\textquotedblright\ SP). It is evident that the RF decreases the SP $S_{0}$ by a factor
of $J_{0}^{2}(\zeta )<1$. Finally, we would like to mention that the relativistic effects in Eq.~\eqref{eq:28} are
given by the second term in the square brackets and by the dispersion function $D(k,\omega )$. In a nonrelativistic
limit ($\beta\to 0$) the former vanishes while the latter is replaced by $D(k,\omega )\to k^{2}$.

To illustrate the features of the stopping of the relativistic charged particles in a laser-irradiated plasma,
we consider below two examples when the intensity of the RF is small but the angle between the vectors $\mathbf{u}$
and $\mathbf{a}$ is arbitrary (Sec.~\ref{sec:3-1}). In the second example we assume that the polarization vector
$\mathbf{E}_{0}$ of the RF is parallel to the beam velocity $\mathbf{u}$ (Sec.~\ref{sec:3-2}).

\subsection{Stopping power at weak intensities of the RF}
\label{sec:3-1}

In this section we consider the case of a weak radiation field ($v_{E}\ll u$) at arbitrary angle $\vartheta$ between the
velocity $\mathbf{u}$ of the beam and the direction of polarization of the RF $\mathbf{a}$. Assuming a point-like test
particle (with $Q(\mathbf{k})=1$), in Eq.~\eqref{eq:28} we keep only the quadratic terms with respect to the quantity
$\mathbf{a}$ (or $\mathbf{v}_{E}$) and for the stopping power $S$ we obtain
\begin{eqnarray}
&&S = S_{0}+\frac{Z^{2}e^{2}}{8\pi u}\int_{-\infty }^{\infty }\omega d\omega
\int \delta (\omega -\mathbf{k}\cdot \mathbf{u})(\mathbf{k}\cdot \mathbf{a})d%
\mathbf{k}  \nonumber \\
&&\times \left\{ \frac{|\Omega _{1}|}{\Omega _{1}}\frac{\delta \left(
\varepsilon (\Omega _{1})\right) }{D(k,\omega )}\left[ (\mathbf{k}\cdot
\mathbf{a})+\frac{\omega _{p}^{2}}{\Omega _{1}^{2}}\frac{(\mathbf{u}\cdot
\mathbf{v}_{E})}{c^{2}}\right] \right.  \label{eq:31} \\
&&+\frac{|\Omega _{-1}|}{\Omega _{-1}}\frac{\delta \left( \varepsilon
(\Omega _{-1})\right) }{D(k,\omega )}\left[ (\mathbf{k}\cdot \mathbf{a})-%
\frac{\omega _{p}^{2}}{\Omega _{-1}^{2}}\frac{(\mathbf{u}\cdot \mathbf{v}%
_{E})}{c^{2}\zeta }\right]  \nonumber \\
&&-\frac{2|\omega |}{\omega }\frac{(\mathbf{k}\cdot \mathbf{a})}{k^{2}%
}\delta \left( \varepsilon (\omega )\right)\Bigg\} ,  \nonumber
\end{eqnarray}%
where $\Omega_{\pm 1}=\omega \pm \omega _{0}$ and $S_{0}$ is the field-free SP given by Eq.~\eqref{eq:30}. The remaining
terms are proportional to the intensity of the radiation field ($\sim a^{2}$). Note that due to the isotropy of the
dielectric function $\varepsilon (\omega )$ as well as the dispersion function $D(k,\omega )$ the angular integrations
in Eq.~\eqref{eq:31} can be easily done. This results in
\begin{equation}
S=S_{0}+\frac{Z^{2}e^{2}\omega _{p}^{2}\alpha ^{2}}{16u^{2}}\left[ \Phi
_{1}(\nu )\left( 3\cos ^{2}\vartheta -1\right) +2\beta ^{2}\Phi _{2}(\nu
)\cos ^{2}\vartheta \right] ,
\label{eq:32}
\end{equation}%
where $\alpha =v_{E}/u$ is the scaled quiver velocity and two functions have been introduced as follows
\begin{eqnarray}
&&\Phi _{1}(\nu )=\frac{\zeta _{+}}{\nu }\left( \gamma ^{-2}\zeta
_{+}^{2}+\beta ^{2}\nu ^{2}\right) \ln \left( 1+\frac{\eta ^{2} \nu^{2}}{\gamma
^{-2}\zeta _{+}^{2}+\beta ^{2}\nu ^{2}}\right)  \label{eq:33} \\
&&-\frac{\zeta _{-}}{\nu }\left( \gamma ^{-2}\zeta _{-}^{2}+\beta ^{2}\nu
^{2}\right) \ln \left( 1+\frac{\eta ^{2} \nu^{2}}{\gamma ^{-2}\zeta _{-}^{2}+\beta
^{2}\nu ^{2}}\right) -2\nu ^{2}\ln (1+ \eta ^{2}),  \nonumber \\
&&\Phi _{2}(\nu )=\zeta _{+}\ln \left( 1+\frac{\eta ^{2} \nu^{2}}{\gamma ^{-2}\zeta
_{+}^{2}+\beta ^{2}\nu ^{2}}\right) +\zeta _{-}\ln \left( 1+\frac{\eta ^{2} \nu^{2}}{%
\gamma ^{-2}\zeta _{-}^{2}+\beta ^{2}\nu ^{2}}\right)  \label{eq:34}
\end{eqnarray}%
with $\eta =k_{\max}u/\omega _{p}$, $\nu =\omega _{p}/\omega _{0}<1$, $\zeta _{\pm }=1\pm \nu $. Also $\gamma $ is the
relativistic factor of the beam with $\gamma ^{-2}=1-\beta ^{2}$. In particular, in the case of the ultrarelativistic
beam with $\gamma \gg \omega_{0}/\omega_{p}$ from Eqs.~\eqref{eq:33} and \eqref{eq:34} we obtain
\begin{eqnarray}
&&\Phi _{1}(\nu )=\frac{6}{\gamma^{2}}\left[\ln ( 1+ \eta ^{2}) -
\frac{\eta ^{2}}{\eta ^{2} +1} \right] ,  \label{eq:33a} \\
&&\Phi _{2}(\nu )=2\ln ( 1+ \eta ^{2}) ,  \label{eq:34a}
\end{eqnarray}%
and in Eq.~\eqref{eq:32} the additional energy loss of the beam (i.e. the term of Eq.~\eqref{eq:32} proportional to
$\sim \alpha^{2}$) is mainly determined by the function $\Phi _{2}(\nu )$.

\begin{figure}[tbp]
\centering{
\includegraphics[width=70mm]{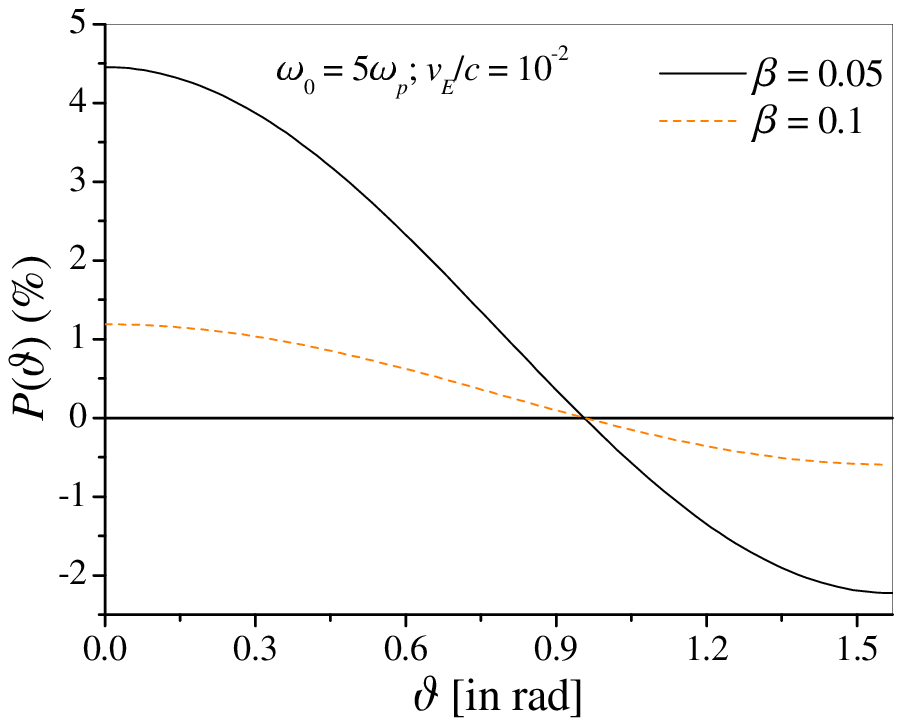}
\includegraphics[width=75mm]{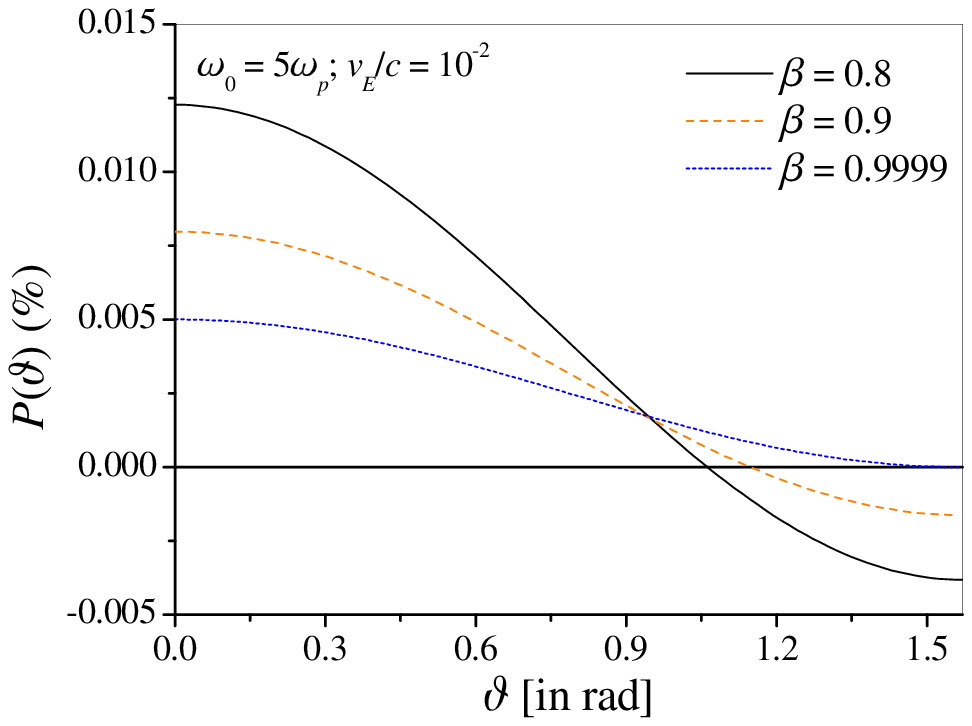}}
\caption{The dimensionless quantity $P(\vartheta )$ (in percents) vs the angle $\vartheta$ (in rad) for the intensity
parameter of the laser field $v_{E}/c =10^{-2}$ and for $\omega _{0}=5\omega_{p}$. Left and right panels demonstrate
the quantity $P(\vartheta )$ for the stopping of the nonrelativistic and relativistic charged particles, respectively,
at various values of the relativistic factor $\beta$. \textbf{Left panel}, $\beta =0.05$ (solid line) and $\beta =0.1$
(dashed line). \textbf{Right panel}, $\beta =0.8$ (solid line), $\beta =0.9$ (dashed line), and $\beta =0.9999$ (dotted
line).}
\label{fig:1}
\end{figure}

Consider next the angular distribution of the SP at low-intensities of the RF. Figure~\ref{fig:1} shows the dimensionless
quantity $P(\vartheta )=[S(\vartheta )-S_{0}]/S_{0}$ (the relative deviation of $S$ from $S_{0}$) as a function of the
angle $\vartheta $ for the scaled laser intensity $v_{E}/c=10^{-2}$ and for $\omega _{0}=5\omega _{p}$, $Z=1$, $A=c/\omega
_{p}r_{e}=1.88\times 10^{7}$, where $r_{e}=e^{2}/mc^{2}$ is the electron classical radius. The quantity $A$ corresponds
to the density of the plasma $n_{e}=10^{22}$ cm$^{-3}$. In addition, left and right panels of Fig.~\ref{fig:1} demonstrate
$P(\vartheta )$ for the stopping of the nonrelativistic (with $\beta =0.05$ and $\beta =0.1$) and relativistic (with
$\beta =0.8$, $\beta =0.9$ and $\beta =0.9999$) charged particles, respectively. At nonrelativistic velocities the second
term in Eq.~\eqref{eq:32} can be neglected and the angular distribution of the quantity $P(\vartheta )$ has a quadrupole
nature (the angular average of $P(\vartheta )$ vanishes). At $0\leqslant\vartheta \leqslant\vartheta_{0}=\arccos (1/\sqrt{3})$
the excitation of the waves with the frequencies $\omega_{0}\pm \omega_{p}$ leads to additional energy loss. At
$\vartheta_{0}\leqslant \vartheta \leqslant \pi /2$ the energy loss changes the sign and the total energy loss decreases. When
the particle moves at $\vartheta =\vartheta_{0}$ with respect to the polarization vector $\mathbf{a}$ the radiation field
has no any influence on the SP. At the relativistic velocities (Fig.~\ref{fig:1}, right panel) the field-free SP $S_{0}$
as well as the relative deviation $P(\vartheta )$ are strongly reduced and the critical angle $\vartheta_{0}$, which now
depends on the factor $\beta$, is shifted towards higher values. Finally, at ultrarelativistic velocities (Fig.~\ref{fig:1},
dotted line in the right panel), $P(\vartheta )$ is positive for arbitrary $\vartheta$ and the radiation field systematically
increases the energy loss of the particle (see also Eq.~\eqref{eq:32} with Eqs.~\eqref{eq:33a} and \eqref{eq:34a}).

\subsection{Stopping power at parallel ($\mathbf{u}\parallel \mathbf{a}$) configuration}
\label{sec:3-2}

Let us now investigate the influence of the radiation field on the stopping process of the relativistic charged particle when
its velocity $\mathbf{u}$ is parallel to $\mathbf{a}$. It is expected that the effect of the RF is maximal in this case. In
the case of the point-like particles after straightforward integrations in Eq.~\eqref{eq:28} we arrive at
\begin{equation}
S=\frac{Z^{2}e^{2}\omega _{p}\omega _{0}}{2u^{2}}\sum_{\ell =-\infty
}^{\infty }\left( \nu +\ell \gamma ^{-2}\right) J_{\ell }^{2}\left( \alpha
(\ell +\nu )\right) \ln \left[ 1+\frac{\eta ^{2} \nu^{2}}{\gamma ^{-2}(\ell +\nu
)^{2}+\beta ^{2}\nu ^{2}}\right] .
\label{eq:35}
\end{equation}
We have used here the same notations as in Sec.~\ref{sec:3-1}. From Eq.~\eqref{eq:35} it is seen that ''no photon'' SP (the
term with $\ell =0$) oscillates with the intensity of the laser field. However, the radiation field suppresses the excitation
of the plasma modes and this term by a factor of $J_{0}^{2}(\alpha \nu )$ is smaller than the field-free SP $S_{0}$. As follows
from Eq.~\eqref{eq:35} at high-intensities of the RF the \textquotedblleft no photon\textquotedblright\ SP vanishes when
$\alpha \nu =(v_{E}/u)(\omega_{p}/\omega _{0})\simeq \mu _{n}$ with $n=1,2,\ldots$, where $\mu _{n}$ are the zeros of the Bessel
function $J_{0}(\mu _{n})=0$ ($\mu_{1}=2.4$, $\mu_{2}=5.52$, \ldots). Then the energy loss of the charged particle is mainly
determined by other terms in Eq.~\eqref{eq:35} and is stipulated by excitation of plasma waves with the combinational frequencies
$\ell \omega_{0}\pm \omega_{p}$.

\begin{figure}[tbp]
\centering{
\includegraphics[width=70mm]{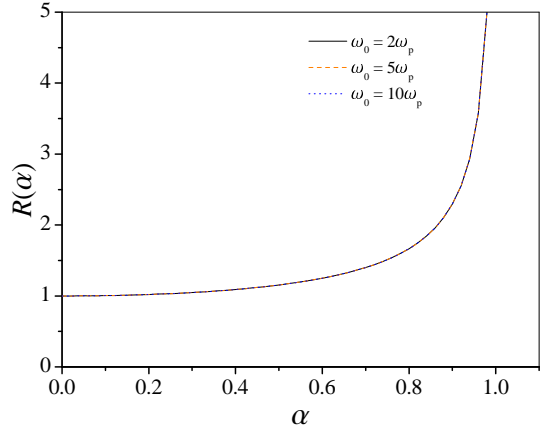}}
\caption{The ratio $R(\alpha )=S(\alpha )/S_{0}$ for an ultrarelativistic projectile particle as a function of
dimensionless intensity of the RF $\alpha$ at $\omega_{0}= 2\omega_{p}$ (solid line), $\omega_{0} =5\omega_{p}$
(dashed line), $\omega_{0} =10\omega_{p}$ (dotted line).}
\label{fig:inf}
\end{figure}

In the case of the ultrarelativistic charged particle (with $\gamma \gg \omega_{0}/\omega_{p}$) Eq.~\eqref{eq:35} gets essentially
simplified. In this equation neglecting the small corrections proportional to $\gamma^{-2}$ we obtain
\begin{equation}
S=\frac{Z^{2}e^{2}\omega _{p}^{2}}{2c^{2}}\ln (1+ \eta ^{2})
\sum_{\ell =-\infty }^{\infty }J_{\ell }^{2}\left( \alpha (\ell +\nu )\right)
\label{eq:36}
\end{equation}%
with $\alpha =v_{E}/c$. Equation~\eqref{eq:36} can be further simplified neglecting the small quantity $\nu =\omega_{p}/\omega_{0}$
in the argument of the Bessel function. The remaining infinite sum is evaluated in Appendix~\ref{sec:app1} [see Eq.~\eqref{eq:ap3}]
which results in
\begin{equation}
S\simeq \frac{Z^{2}e^{2}\omega _{p}^{2}}{2c^{2}}\ln (1+ \eta ^{2}) \frac{1}{\sqrt{1-\alpha ^{2}}} .
\label{eq:37}
\end{equation}%
Thus, in the case of the ultrarelativistic charged particles the energy loss of the beam essentially increases with the
intensity of the RF. Let us recall, however, that Eq.~\eqref{eq:37} is only valid below relativistic intensities of the
RF when $\alpha\lesssim 1$. For treating precisely the resonance in Eq.~\eqref{eq:37} at $\alpha =1$ we should go beyond
the present long-wavelength approximation of the laser field. Figure~\ref{fig:inf} demonstrates the quantity $R(\alpha )=
S(\alpha )/S_{0}$ for three distinct values of the laser frequency $\omega_{0}$ as a function of the RF intensity
$\alpha =v_{E}/c$, where the SP is given by Eq.~\eqref{eq:36} for an
ultrarelativistic projectile particle. Note that the quantity $R(\alpha )$ is independent on the plasma density $n_{e}$.
It is seen that the SP is indeed not sensitive to the variation of the laser frequency and the approximation~\eqref{eq:37}
is justified at $\alpha\lesssim 1$.

Next let us examine an interesting regime of the high-intensities of the RF when $\alpha \nu >1$. It is clear that such
a regime requires nonrelativistic velocities of the charged particles beam. Thus, at $\alpha\nu >1$ we consider here the
relativistic effects as small corrections to the corresponding nonrelativistic expressions. Using the asymptotic behavior
of the Bessel function \citep{gra80} in a general Eq.~\eqref{eq:35} it is straightforward to obtain
\begin{eqnarray}
&&S = \frac{Z^{2}e^{2}\omega _{p}^{2}}{u^{2}}\left\{ \frac{1}{2}%
J_{0}^{2}(\alpha \nu )\ln (1+ \eta ^{2})+\frac{1}{\pi \gamma ^{2}}%
\frac{\sin (2\alpha \nu )}{\alpha \nu }\Gamma _{1}(\eta ,\alpha )\right. \label{eq:38} \\
&&\left. +\frac{\beta ^{2}}{\pi \alpha }\left[ \cos (2\alpha \nu )\Gamma
_{2}(\eta ,\alpha )+\Gamma (\eta )\right] \right\} ,  \nonumber
\end{eqnarray}%
where we have introduced the following quantities
\begin{eqnarray}
&&\Gamma _{1}(\eta ,\alpha ) = \sum_{\ell =1}^{\infty }(-1)^{\ell }\ln \left(
1+\frac{\eta ^{2} \nu^{2}}{\gamma ^{-2}\ell ^{2}+\beta ^{2}\nu ^{2}}\right) \cos
(2\alpha \ell ),  \label{eq:39} \\
&&\Gamma _{2}(\eta ,\alpha ) = \sum_{\ell =1}^{\infty }\frac{(-1)^{\ell }}{%
\ell }\ln \left( 1+\frac{\eta ^{2} \nu^{2}}{\gamma ^{-2}\ell ^{2}+\beta ^{2}\nu ^{2}}%
\right) \sin (2\alpha \ell ),  \label{eq:40} \\
&&\Gamma (\eta ) = \sum_{\ell =1}^{\infty }\frac{1}{\ell }\ln \left( 1+\frac{%
\eta ^{2} \nu^{2}}{\gamma ^{-2}\ell ^{2}+\beta ^{2}\nu ^{2}}\right) .  \label{eq:41}
\end{eqnarray}%
In Appendix~\ref{sec:app1} we have found some approximate analytical expressions for these functions. Recalling that
$\gamma^{-2}=1-\beta ^{2}$ the relativistic corrections ($\sim \beta^{2}$) can be easily deduced from Eq.~\eqref{eq:38}.

Let us consider briefly two particular cases. When $\alpha =\pi (s+\frac{1}{2})$ (where $s=0,1,\ldots$) all terms in
Eq.~\eqref{eq:39} have a positive sign and the function $\Gamma_{1}(\eta ,\alpha)$ in Eq.~\eqref{eq:38} is maximal in
this case due to the constructive interference of the excited waves with the frequencies $\ell \omega_{0}\pm \omega_{p}$.
Note that the relativistic correction given by $\Gamma_{2}(\eta ,\alpha )$ vanishes in this case. From Eq.~\eqref{eq:ap6}
it follows that $p(\alpha )=\frac{1}{2}$ and Eq.~\eqref{eq:ap7} can be evaluated explicitly. The result reads
\begin{equation}
\Gamma _{1}(\eta ,\alpha )=\ln \left\{ \frac{A(0)}{A(\eta )}\frac{\sinh [\pi
\gamma A(\eta )]}{\sinh [\pi \gamma A(0)]}\right\} ,
\label{eq:41a}
\end{equation}%
where $A(\eta )=\nu\sqrt{\beta^{2} +\eta^{2}}$, $A(0)=\beta \nu $.

In the second regime when $\alpha =\pi (s+1)$, $p(\alpha )=0$ and the relativistic corrections given by $\Gamma_{2}$ again
vanishes, $\Gamma_{2}(\eta ,\alpha )=0$. Equation~\eqref{eq:ap7} then yields a negative contribution to the energy loss
\eqref{eq:38} with
\begin{equation}
\Gamma _{1}(\eta ,\alpha )=-\ln \left\{ \frac{A(\eta )}{A(0 )}\frac{\tanh [\frac{%
1}{2}\pi \gamma A(0 )]}{\tanh [\frac{1}{2}\pi \gamma A(\eta )]}\right\} <0 .
\label{eq:41b}
\end{equation}%
The function $\Gamma_{1}(\eta ,\alpha )$ is minimal in this case since the different harmonics cancel effectively each
other. However, it should be emphasized that the contribution of the term containing the function $\Gamma_{1}(\eta ,\alpha )$
to the SP \eqref{eq:38} can either be positive or negative depending on the dimensionless quantity $\alpha\nu$. Moreover,
the absolute value of this contribution decreases with the parameter $\alpha\nu$.

\begin{figure}[tbp]
\centering{
\includegraphics[width=70mm]{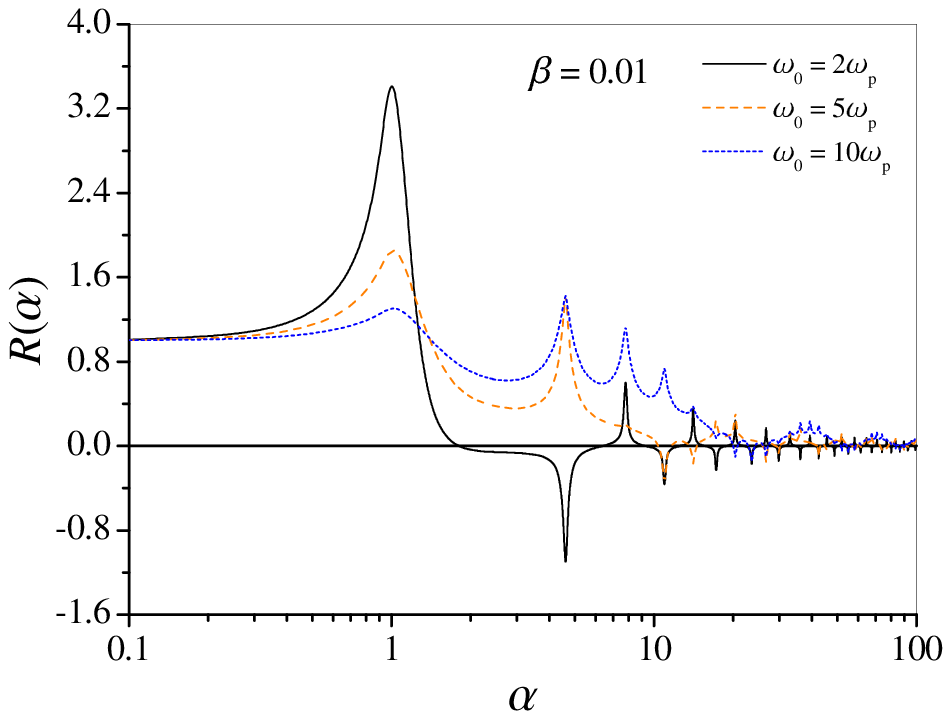}
\includegraphics[width=70mm]{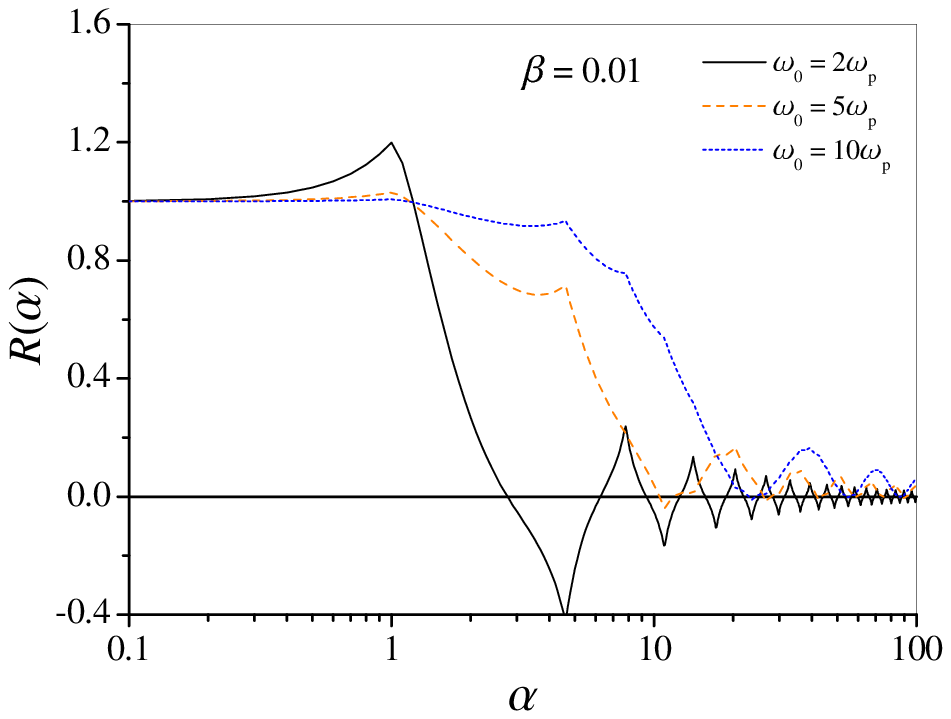}
}
\caption{(\textbf{Left panel}) the ratio $R(\alpha )=S(\alpha )/S_{0}$
as a function of dimensionless quantity $\alpha $ at $n_{e} =10^{22}$~cm$^{-3}$, $\beta =10^{-2}$, $\omega_{0}= 2\omega_{p}$ (solid line),
$\omega_{0} =5\omega_{p}$ (dashed line), $\omega_{0} =10\omega_{p}$ (dotted line).
(\textbf{Right panel}) same as in the left panel but for $n_{e} =10^{24}$~cm$^{-3}$.}
\label{fig:2}
\end{figure}

\begin{figure}[tbp]
\centering{
\includegraphics[width=70mm]{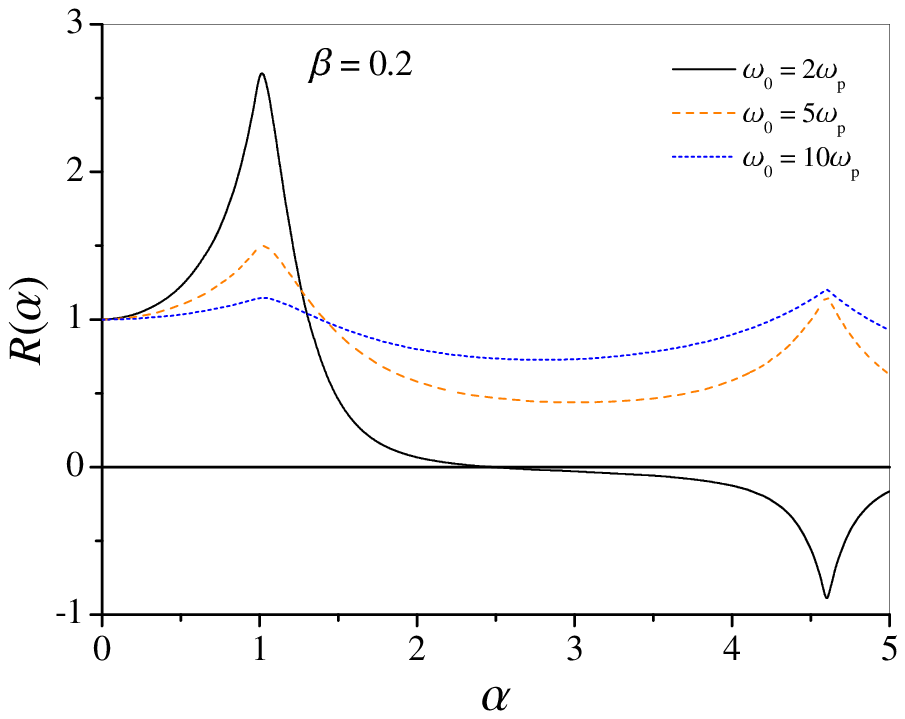}
\includegraphics[width=70mm]{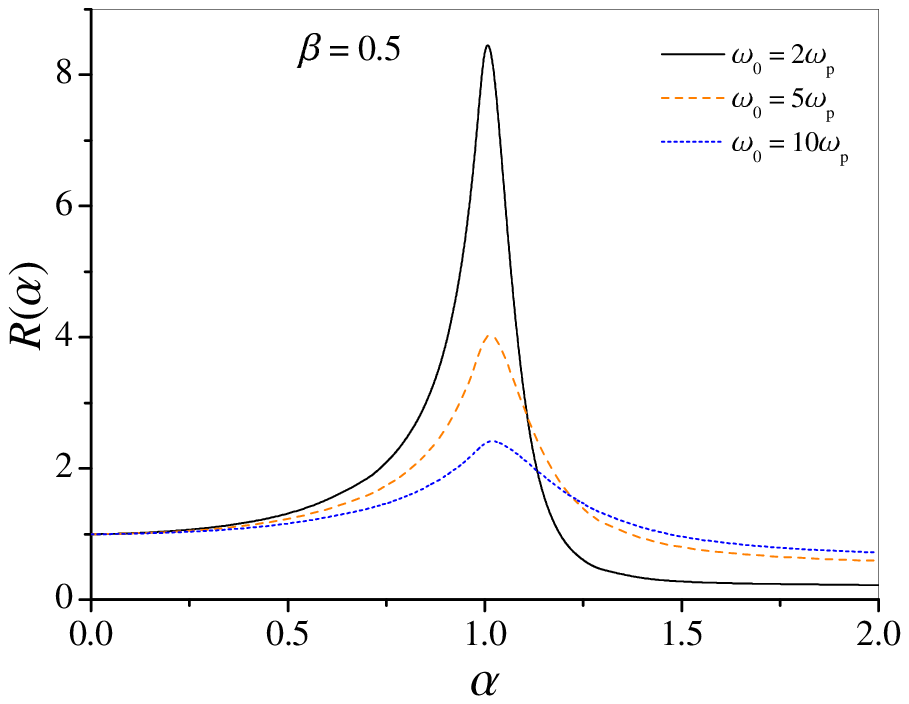}
}
\caption{(\textbf{Left panel}) the ratio $R(\alpha )=S(\alpha )/S_{0}$
as a function of dimensionless quantity $\alpha $ at $\beta =0.2$, $\omega_{0}= 2\omega_{p}$ (solid line),
$\omega_{0} =5\omega_{p}$ (dashed line), $\omega_{0} =10\omega_{p}$ (dotted line).
(\textbf{Right panel}) same as in the left panel but for $\beta =0.5$.}
\label{fig:3}
\end{figure}

\begin{figure}[tbp]
\centering{
\includegraphics[width=70mm]{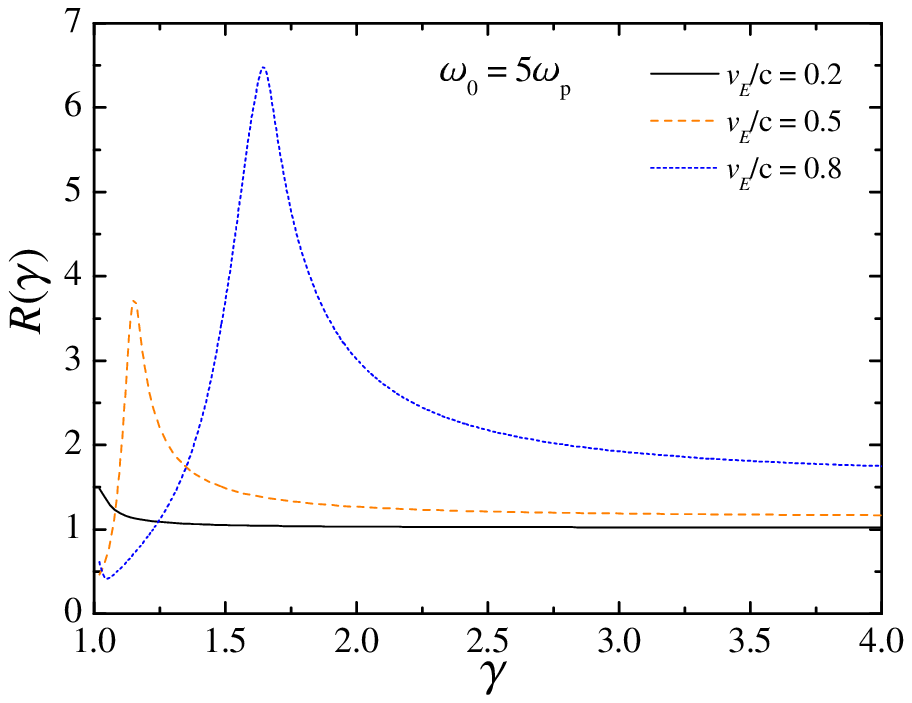}
\includegraphics[width=70mm]{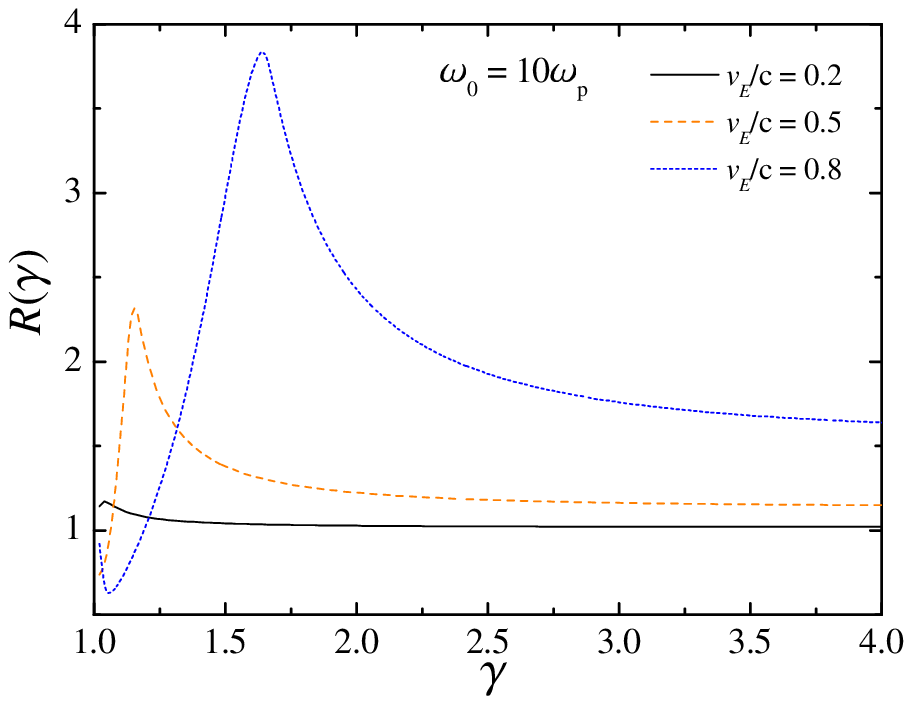}
}
\caption{(\textbf{Left panel}) the ratio $R(\gamma )=S(\gamma )/S_{0}$ as a function of the relativistic
factor $\gamma$ of the electron beam at $\omega_{0} =5\omega_{p}$, $v_{E}/c =0.2$ (solid line), $v_{E}/c =0.5$
(dashed line), $v_{E}/c =0.8$ (dotted line). (\textbf{Right panel}) same as in the left panel but for
$\omega_{0} =10\omega_{p}$.}
\label{fig:4}
\end{figure}

The results of the numerical evaluation of the SP given by a general Eq.~\eqref{eq:35} are shown in Figs.~\ref{fig:2}--\ref{fig:4}
both for nonrelativistic and relativistic projectiles, where the ratios $R(\alpha )=S(\alpha )/S_{0}$ and $R(\gamma )
=S(\gamma )/S_{0}$ are plotted as the functions of the RF intensity $\alpha =v_{E}/u=0.855W_{L}^{1/2}\lambda_{0}/\beta$
(Figs.~\ref{fig:2} and \ref{fig:3}) and the relativistic factor $\gamma$ of the electron beam (Fig.~\ref{fig:4}), respectively.
Here $W_{L}$ and $\lambda_{0}$ are measured in units $10^{18}$~W/cm$^{2}$ and $\mu$m, respectively. Also in Figs.~\ref{fig:2}
and \ref{fig:3} the parameter $\alpha$ varies in the interval $0\lesssim\alpha <1/\beta$ to ensure the applicability of the
long-wavelength approximation of the laser field. From Fig.~\ref{fig:2} it is seen that the SP of a nonrelativistic particle
exhibits a strong oscillations. Furthermore, it may exceed the field-free SP and change the sign due to plasma irradiation
by intense ($\alpha >1$) laser field. These properties of a nonrelativistic SP have been obtained previously for a classical
plasma \citep{ner99} as well as for a fully degenerated plasma \citep{ner11}. The effect of the acceleration of the projectile
particle shown in Fig.~\ref{fig:2} occurs at $v_{E}/u\simeq \mu_{\ell}/\nu$ (with $\ell =1,2,\ldots $) when the ''no photon''
SP nearly vanishes. It should be noted that in the laser irradiated plasma a parametric instability is expected \citep{ner12}
with an increment increasing with the intensity of the radiation field. This restricts the possible acceleration time of the
incident particle. The similar behavior of the SP is demonstrated in the right panel of Fig.~\ref{fig:2}, where however the
plasma density is $n_{e}=10^{24}$~cm$^{-3}$ and two orders of magnitude exceeds the value adopted in the left panel of
Fig.~\ref{fig:2}. We note that although the ratio $R(\alpha )$ is reduced with a plasma density, the field-free SP given by
Eq.~\eqref{eq:30} is proportional to $n_{e}$ which in turn results in a strong enhancement of the SP $S(\alpha) =S_{0}R(\alpha)$
with a plasma density.

Next, in Fig.~\ref{fig:3}, we consider weakly relativistic regimes of the electron beam with $\beta =0.2$ (left panel) and
$\beta =0.5$ (right panel). At smaller intensities of the RF, $0<\alpha \lesssim 1$, the SP is described qualitatively by
Eq.~\eqref{eq:37} and reaches its maximum at $\alpha\simeq 1$ (or at $v_{E}\simeq u$). This is because the electrons
are driven by the laser field in the direction of the beam with nearly same velocity $u$ effectively enhancing the
amplitude of the excited waves. The minimum or two maxima of the SP at $\alpha\simeq 3\pi/2$ in the left panel of
Fig.~\ref{fig:3} are formed due to the constructive interference of the excited harmonics as discussed above (see
Eq.~\eqref{eq:41a}). The other minima of the SP at $\alpha\simeq \pi$ in the left panel of Fig.~\ref{fig:3} are formed
due to the destructive interference of the excited harmonics (see Eq.~\eqref{eq:41b}). Let us note that at $\alpha\sim 1$
the effect of the enhancement of the SP of the test particle moving in a laser irradiated plasma is intensified at smaller
frequency of the laser field $\omega_{0}= 2\omega_{p}$ (see Figs.~\ref{fig:2} and \ref{fig:3}) while at higher intensities,
$\alpha >1$, the quantity $R(\alpha)$ increases with the frequency $\omega_{0}$.

Finally, in Fig.~\ref{fig:4}, the quantity $R(\gamma)$ is shown as a function of the electron beam energy at various
intensities of the laser field given here by a dimensionless parameter $v_{E}/c$. It is seen that at the moderate energies
of the electron beam the SP strongly increases with the laser intensity as has been also demonstrated in Figs.~\ref{fig:2}
and \ref{fig:3}. Comparing two panels of Fig.~\ref{fig:4} we may conclude that the position of the maximum of the SP depends
only on the intensity of the RF, $\gamma_{\max} \sim v_{E}/c$, while the maximum value of the SP is scaled as $R_{\max} \sim
(v_{E}/c)(\omega_{p}/\omega_{0})^{1/2}$ and is sensitive to both the intensity and frequency of the RF. Furthermore, with
increasing the relativistic factor $\gamma$ of the electron beam one asymptotically arrives at the ultrarelativistic regime
described by Eqs.~\eqref{eq:36} and \eqref{eq:37}. Interestingly this occurs at the moderate energies of the electron beam
with $\gamma\gtrsim 4$ which, in particular, pertain to the FIS relativistic electron beam in the typical 1- to 2-MeV energy
range of practical interest.

\section{Conclusions}
\label{sec:4}

In this paper, we have investigated the energy loss of a relativistic charged particles beam moving in a laser irradiated
plasma, where the laser field has been treated in the long-wavelength approximation. The dynamics of the beam-plasma system
in the presence of the RF is studied by the linearized fluid and Maxwell equations for the plasma particles and the
electromagnetic fields, respectively. The full electromagnetic response of the system is derived involving all harmonics
of the RF. It has been shown that, in general, the excited longitudinal and transversal modes are parametrically coupled
due to the presence of the RF. As a result, the \v{C}erenkov radiation by the charged particles beam, which is absent in
a plasma in the field-free case, becomes possible. However, in contrast to the usual \v{C}erenkov effect in a laser-free
medium, the spectrum of the radiation in the present case is discrete and the emitted frequencies are localized around the
harmonics of the RF depending essentially on the energy of the incoming beam. In particular, at ultrarelativistic energies
of the beam emitted frequencies are shifted towards higher harmonics of the RF.

In the course of our study, we have derived a general expression for the SP, which has
been also simplified in the limit of a low-intensity laser field. As in the field-free case, the SP in a laser irradiated
plasma is completely determined by the dielectric function of the plasma which in the present context is given by the
hydrodynamic approximation. Explicit calculations have been done for two particular cases assuming (i) a low-intensity laser
field but arbitrary orientation of the laser polarization vector $\mathbf{E}_{0}$ with respect to the charged particles beam
velocity $\mathbf{u}$, and (ii) an arbitrary (but nonrelativistic) intensity of the laser field when $\mathbf{E}_{0}$ is
parallel to $\mathbf{u}$. In the case (i) it has been shown that the RF increases the SP for small angles between the
velocity of the beam and $\mathbf{E}_{0}$ while decreasing it at large angles. Furthermore, the relative deviation of the
SP from the field-free value is strongly reduced with increasing the beam energy. In the case (ii) and at high-intensities
of the RF two extrem regimes have been distinguished when the excited harmonics cancel effectively each other reducing
strongly the energy loss or increasing it due to the constructive interference. As in the previous nonrelativistic treatments
\citep{ner99,ner11} an acceleration of the projectile particle is expected at high-intensities of the RF when the quiver
velocity of the plasma electrons exceeds the beam velocity $u$. Special attention has been paid to the relativistic effects
of the beam. We have demonstrated that in general the energy loss increases with the beam energy forming a maximum at moderate
relativistic factors $\gamma\lesssim 2$ which, however, are shifted towards higher energies of the beam with the laser intensity.
In addition, the enhancement of the SP is more pronounced when the laser frequency approaches the plasma frequency in agreement
with PIC simulations \citep{hu11}. Finally, it has been also shown that the SP of the ultrarelativistic electron beam increases
systematically with the intensity of the RF exceeding essentially the field-free SP.

Going beyond the present model, which is based on several approximations, we can envisage a number of improvements.
In particular, a simple possibility qualitatively involving relativistic intensities of the laser field is to replace the
quiver velocity $v_{E}\to V_{E}$ by its relativistic counterpart $V_{E}$, where $V_{E}= v_{E}[1+(v_{E}/c)^{2}]^{-1/2}$. The
latter approximation can be easily obtained from a single-particle dynamics in a plane RF. In the course of our study we have
also neglected the \v{C}erenkov radiation which might be important in the total balance of the energy loss process. However, in
a general Eq.~\eqref{eq:22} this effect is well separated by its spectral characteristics and can be treated independently. We
intend to address this issue in our forthcoming investigations.

\section*{Acknowledgments}
The work of H.B.N. has been supported by the State Committee of Science of the Armenian Ministry of Higher
Education and Science (Project No.~13-1C200).

\appendix

\section{Evaluation of the functions introduced in the text}
\label{sec:app1}

In this Appendix we evaluate the infinite sum containing the square of the Bessel functions and involved in
Eq.~\eqref{eq:36}. Also we will derive some analytical expressions for the functions $\Gamma_{1}(\eta ,\alpha )$ and
$\Gamma_{2}(\eta ,\alpha )$ introduced in Sec.~\ref{sec:3-2} [see Eqs.~\eqref{eq:39} and \eqref{eq:40}].

Our starting point is the Neumann's formula \citep{bat53}
\begin{equation}
J_{\ell }^{2}(z)=\frac{2}{\pi }\int_{0}^{\pi /2}J_{2\ell }(2z\cos t)dt .
\label{eq:ap1}
\end{equation}%
Using this expression as well as the known relation \citep{gra80}
\begin{equation}
\sum_{\ell =1}^{\infty }J_{2\ell }(2\ell z)=\frac{z^{2}}{2(1-z^{2})},
\label{eq:ap2}
\end{equation}%
the desired summation can be performed easily
\begin{equation}
\sum_{\ell =1}^{\infty }J_{\ell }^{2}(\alpha \ell )=\frac{1}{\pi }%
\int_{0}^{\pi /2}\frac{dt}{1-\alpha ^{2}\cos ^{2}t}-\frac{1}{2}=\frac{1}{2}%
\left( \frac{1}{\sqrt{1-\alpha ^{2}}}-1\right) .  \label{eq:ap3}
\end{equation}%
The latter formula is valid at $\alpha <1$.

Next we derive approximate but very accurate expressions for the functions $\Gamma_{1}(\eta ,\alpha )$ and
$\Gamma_{2}(\eta ,\alpha )$ which determine the SP in Eq.~\eqref{eq:38}. For that purpose consider the following
two functions
\begin{eqnarray}
&&G_{1}(a,b) =\sum_{\ell=1}^{\infty}(-1)^{\ell}\ln\left(  1+\frac{a^{2}}%
{\ell^{2}+b^{2}}\right)  \cos(2\alpha\ell) , \label{eq:ap4} \\
&&G_{2}(a,b) =\sum_{\ell=1}^{\infty}\frac{(-1)^{\ell}}{\ell}\ln\left(
1+\frac{a^{2}}{\ell^{2}+b^{2}}\right) \sin(2\alpha\ell) , \label{eq:ap5}
\end{eqnarray}
where $a>0$ and $b>0$ are some arbitrary positive variables. Note that $\Gamma_{1;2}(\eta ,\alpha)=G_{1;2}(a,b)$
at $a=\eta\gamma \nu$ and $b= \beta\gamma \nu$. Consider the derivatives of these functions with respect to the
variable $a$. Using the known summation formulas involving the trigonometric functions \citep{gra80} we arrive at
\begin{eqnarray}
&&\frac{\partial}{\partial a}G_{1}(a,b) =2a\sum_{\ell=1}^{\infty}(-1)^{\ell
}\frac{\cos(2\alpha\ell)}{\ell^{2}+u^{2}}
=\frac{\pi a}{u}\left\{  \frac{\cosh[2\pi up(\alpha)]}{\sinh(\pi u)}%
-\frac{1}{\pi u}\right\} , \label{eq:ap4a} \\
&&\frac{\partial}{\partial a}G_{2}(a,b) =2a\sum_{\ell=1}^{\infty}%
\frac{(-1)^{\ell}}{\ell}\frac{\sin(2\alpha\ell)}{\ell^{2}+u^{2}}
=\frac{\pi a}{u^{2}}\cos\left[  \pi q(\alpha)\right]  \left\{  \frac
{\sinh\left[  2\pi up(\alpha)\right]  }{\sinh(\pi u)}-2p(\alpha)\right\}  \label{eq:ap5a}
\end{eqnarray}
with $u=(a^{2}+b^{2})^{1/2}$, $q(\alpha )=E[M(\alpha /\pi ) +1/2] $, and
\begin{equation}
p(\alpha )=\left\{
\begin{array}{cc}
M\left( \frac{\alpha }{\pi }\right) ; & 0\leqslant M\left( \frac{\alpha }{%
\pi }\right) \leqslant \frac{1}{2} \\
1-M\left( \frac{\alpha }{\pi }\right) ; & \frac{1}{2}\leqslant M\left( \frac{%
\alpha }{\pi }\right) <1%
\end{array}%
\right. ,  \label{eq:ap6}
\end{equation}%
where $M(z)$ and $E(z)$ are the fractional and integer parts of $z$, respectively. Consequently, from
Eqs.~\eqref{eq:ap4a} and \eqref{eq:ap5a} we obtain
\begin{eqnarray}
&&G_{1}(a,b) =G_{1}(a_{0},b)+\int_{\pi u_{0}}^{\pi u}\frac{\cosh[2xp(\alpha)]}{\sinh (x) }dx
-\ln\frac{u}{u_{0}} , \label{eq:ap7} \\
&&G_{2}(a,b) =G_{2}(a_{0},b)+\pi\cos [\pi q(\alpha )] \left\{\int_{\pi u_{0}}^{\pi u}
\frac{\sinh [2xp(\alpha )]}{\sinh (x)}\frac{dx}{x} -2p(\alpha ) \ln\frac{u}{u_{0}}\right\} . \label{eq:ap8}
\end{eqnarray}
Here $u_{0}=(a_{0}^{2}+b^{2})^{1/2}$ and $a_{0} >0$ is an arbitrary initial value of $a$. Thus, the infinite sums
in Eqs.~\eqref{eq:39} and \eqref{eq:40} have been represented in the equivalent integral forms with finite integration
limits. Assuming that $b>1$ let us apply Eqs.~\eqref{eq:ap7} and \eqref{eq:ap8} in a particular case with $a_{0}=0$.
Taking into account the initial conditions (see Eqs.~\eqref{eq:ap4} and \eqref{eq:ap5}) $G_{1}(0,b)=G_{2}(0,b)=0$
from Eqs.~\eqref{eq:ap7} and \eqref{eq:ap8} we obtain
\begin{eqnarray}
&&G_{1}(a,b) =-\ln\frac{u}{b} +\int_{\pi b}^{\pi u}\frac{\cosh [2xp(\alpha)]}{\sinh(x)}dx , \label{eq:ap9a}  \\
&&G_{2}(a,b) =\pi\cos[\pi q(\alpha)]\left\{  \int_{\pi b}^{\pi u}\frac{\sinh [2xp(\alpha)]}
{\sinh (x)}\frac{dx}{x} - 2p(\alpha)\ln\frac{u}{b} \right\} .  \label{eq:ap10a}
\end{eqnarray}
For further progress we note that at $b>1$ the hyperbolic functions in Eqs.~\eqref{eq:ap9a} and \eqref{eq:ap10a}
can be replaced by an exponential function, $\sinh (\pi x)\simeq e^{\pi x}/2$, which yield the following
approximations for the functions $G_{1}(a,b)$ and $G_{2}(a,b)$
\begin{eqnarray}
&&G_{1}(a,b) \simeq -\ln\frac{u}{b} +\frac{1}{2Q_{-}(\alpha)}\left\{  \exp\left[  -2\pi bQ_{-}(\alpha)\right]
-\exp\left[  -2\pi u Q_{-}(\alpha)\right]  \right\}  \label{eq:ap9}  \\
&&+\frac{1}{2Q_{+}(\alpha)}\left\{\exp\left[  -2\pi bQ_{+}(\alpha)\right]
-\exp\left[-2\pi u Q_{+}(\alpha)\right]  \right\} ,  \nonumber  \\
&&G_{2}(a,b) \simeq\pi\cos[\pi q(\alpha)]\Big\{ E_{1}\left[2\pi u
Q_{+}(\alpha) \right] -E_{1}\left[2\pi u Q_{-}(\alpha) \right]  \label{eq:ap10}  \\
&& +E_{1}\left[  2\pi bQ_{-}(\alpha)\right]  -E_{1}\left[  2\pi
bQ_{+}(\alpha)\right] -2p(\alpha)\ln\frac{u}{b} \Big\} , \nonumber
\end{eqnarray}
where $E_{1}(z)$ is the exponential integral function and $Q_{\pm}(\alpha ) =\frac{1}{2}\pm p(\alpha )$.

Let us now consider the opposite case of small $b<1$. Then at $\ell \geqslant 2$ we may neglect
the variable $b$ in Eqs.~\eqref{eq:ap4} and \eqref{eq:ap5} and represent these expressions in the
approximate form
\begin{eqnarray}
&&G_{1}(a,b) \simeq G_{1}(a,0)+ \ln\left(1+\frac{a^{2} b^{2}}{1+a^{2}+b^{2}}\right) \cos(2\alpha)  \label{eq:ap11a}  \\
&&-\ln\left(1+\frac{a^{2} b^{2}/4}{4+a^{2}+b^{2}}\right) \cos(4\alpha) ,  \nonumber  \\
&&G_{2}(a,b) \simeq G_{2}(a,0)+ \ln\left(1+\frac{a^{2} b^{2}}{1+a^{2}+b^{2}}\right) \sin(2\alpha)  \label{eq:ap12a}  \\
&&-\frac{1}{2}\ln\left(1+\frac{a^{2} b^{2}/4}{4+a^{2}+b^{2}}\right) \sin(4\alpha) . \nonumber
\end{eqnarray}
Here $G_{1;2}(a,0)$ are the functions $G_{1;2}(a,b)$ at $b=0$ and are evaluated in a similar way as
approximate Eqs.~\eqref{eq:ap9} and \eqref{eq:ap10}. For that purpose consider again Eqs.~\eqref{eq:ap7}
and \eqref{eq:ap8} with $a_{0}=1$ and $b=0$, that is
\begin{eqnarray}
&&G_{1}(a,0) =G_{1}(1,0)+\int_{\pi}^{\pi a}\frac{\cosh[2xp(\alpha)]}{\sinh (x)}dx -\ln a , \label{eq:ap11}  \\
&&G_{2}(a,0) =G_{2}(1,0)+\pi\cos [\pi q(\alpha )] \left\{\int_{\pi}^{\pi a}\frac{\sinh [2xp(\alpha)]}
{\sinh (x)} \frac{dx}{x} -2p(\alpha ) \ln a\right\} . \label{eq:ap12}
\end{eqnarray}
In the latter expressions replacing the hyperbolic sine functions by the exponential functions we arrive at
the following approximate formulas for the quantities $G_{1}(a,0)$ and $G_{2}(a,0)$
\begin{eqnarray}
&&G_{1}(a,0) \simeq G_{1}(1,0)+\frac{1}{2Q_{-}(\alpha )}\left\{\exp\left[-2\pi Q_{-}(\alpha )\right]
-\exp\left[-2\pi aQ_{-}(\alpha )\right]  \right\}   \label{eq:ap13}   \\
&& +\frac{1}{2Q_{+}(\alpha )}\left\{\exp\left[-2\pi Q_{+}(\alpha ) \right] -\exp\left[-2\pi aQ_{+}(\alpha )
\right]  \right\}  -\ln a ,   \nonumber  \\
&&G_{2}(a,0) \simeq G_{2}(1,0)+\pi\cos\left[\pi q(\alpha )\right] \left\{E_{1}\left[2\pi Q_{-}(\alpha )\right]
-E_{1}\left[2\pi Q_{+}(\alpha ) \right]  \right.   \label{eq:ap14}  \\
&& \left.  +E_{1}\left[2\pi aQ_{+}(\alpha ) \right] -E_{1}\left[2\pi aQ_{-}(\alpha )\right] -2p(\alpha )
\ln a\right\} .  \nonumber
\end{eqnarray}

Finally, for the quantities $G_{1}(1,0)$ and $G_{2}(1,0)$ it follows from Eqs.~\eqref{eq:ap4}
and \eqref{eq:ap5} that \citep{gra80}
\begin{eqnarray}
&&G_{1}(1,0)\simeq C_{1} \cos (2\alpha ) -C_{2} \cos (4\alpha ) +\pi ^{2}\left[ p^{2}(\alpha )-\frac{1}{12}\right] ,  \label{eq:ap15} \\
&&G_{2}(1,0) \simeq C_{1} \sin (2\alpha ) -\frac{1}{2} C_{2}\sin (4\alpha )
+\frac{2}{3}\pi ^{3}\cos \left[ \pi q(\alpha )\right] p(\alpha )\left[ p^{2}(\alpha )-\frac{1}{4}\right] ,  \label{eq:ap16}
\end{eqnarray}%
where $C_{1} =1-\ln 2$, $C_{2} =1/4 - \ln (5/4)$.
For approximate evaluation of the quantities $G_{1}(1,0)$ and $G_{2}(1,0)$ we have used the expansion $\ln (1+\ell^{-2})\simeq \ell^{-2}$
at $\ell \geqslant 3$ for the logarithmic function. Then at $b<1$ the final analytical expressions for $G_{1}(a,b)$
and $G_{2}(a,b)$ are obtained from Eqs.~\eqref{eq:ap11a}, \eqref{eq:ap13}, \eqref{eq:ap15} and \eqref{eq:ap12a},
\eqref{eq:ap14}, \eqref{eq:ap16}, respectively. The relative accuracy of the derived approximations is less than $10^{-3}$
in a wide range of the parameters.


\begin{thebibliography}{99}
\bibitem[Akopyan \emph{et al.}, 1997]{ako97} \textsc{Akopyan}, E.A., \textsc{Nersisyan}, H.B. \& \textsc{Matevosyan}, H.H. (1997).
Energy losses of a charged particle in a plasma in an external field allowing for the field action on plasma and particle motion.
\emph{Radiophys. Quant. Electrons.} \textbf{40}, 823--826.


\bibitem[Alexandrov \emph{et al.}, 1984]{ale84} \textsc{Alexandrov}, A.F., \textsc{Bogdankevich}, L.S.
\& \textsc{Rukhadze}, A.A. (1984).
\emph{Principles of Plasma Electrodynamics.} Heidelberg: Springer.


\bibitem[Aliev \emph{et al.}, 1971]{ali71} \textsc{Aliev}, Yu.M., \textsc{Gorbunov}, L.M. \& \textsc{Ramazashvili}, R.R. (1971).
Polarization losses of a fast heavy particle in a plasma located in a strong high frequency field.
\emph{Zh. Eksp. Teor. Fiz.} \textbf{61}, 1477--1480.


\bibitem[Arista \emph{et al.}, 1989]{ari89} \textsc{Arista}, N.R., \textsc{Galv\~{a}o} R.O.M. \& \textsc{Miranda}, L.C.M. (1989).
Laser-field effects on the interaction of charged particles with a degenerate electron gas.
\emph{Phys. Rev. A} \textbf{40}, 3808--3816.


\bibitem[Arista \emph{et al.}, 1990]{ari90} \textsc{Arista}, N.R., \textsc{Galv\~{a}o} R.O.M. \& \textsc{Miranda}, L.C.M. (1990).
Influence of a strong laser field on the stopping power for charged test particles in nondegenerate plasmas.
\emph{J. Phys. Soc. Jpn.} \textbf{59}, 544--552.


\bibitem[Bateman \& Erd\'{e}lyi, 1953]{bat53} \textsc{Bateman,} H. \& \textsc{Erdelyi}, A. (1953).
\emph{Higher Transcendental Functions}, vol. 2. New York: McGraw-Hill.


\bibitem[Couillaud \emph{et al.}, 1994]{cou94} \textsc{Couillaud}, C.,
\textsc{Deicas}, R., \textsc{Nardin}, Ph., \textsc{Beuve}, M.A., \textsc{%
Guihaume}, J.M., \textsc{Renaud}, M., \textsc{Cukier}, M., \textsc{Deutsch},
C. \& \textsc{Maynard}, G. (1994). Ionization and stopping of heavy ions in
dense laser--ablated plasmas. \emph{Phys. Rev. E} \textbf{49}, 1545--1562.


\bibitem[D'Avanzo \emph{et al.}, 1993]{ava93} \textsc{D'Avanzo}, J., \textsc{%
Lontano}, M. \& \textsc{Bortignon}, P.F. (1993). Fast--ion interaction in
dense plasmas with two--ion correlation effects. \emph{Phys. Rev. E} \textbf{%
47}, 3574--3584.


\bibitem[Deutsch, 1986]{deu86} \textsc{Deutsch}, C. (1986).
Inertial confinement fusion driven by intense ion beams.
\emph{Ann. Phys. Paris} \textbf{11}, 1--111.


\bibitem[Deutsch, 1995]{deu95} \textsc{Deutsch}, C. (1995).
Correlated stopping of Coulomb clusters in a dense jellium target.
\emph{Phys. Rev. E} \textbf{51}, 619--631.


\bibitem[Deutsch \& Fromy, 2001]{deu01} \textsc{Deutsch}, C. \& \textsc{Fromy}, P. (2001).
Correlated stopping of relativistic electron beams in supercompressed DT fuel.
\emph{Nucl. Instrum. Methods Phys. Res. A} \textbf{464}, 243--246.


\bibitem[Deutsch \emph{et al.}, 1996]{deu96} \textsc{Deutsch}, C., \textsc{Furukawa},
H., \textsc{Mima}, K., \textsc{Murakami}, M. \& \textsc{Nishihara}, K. (1996). Interaction
physics of the fast ignitor concept. \emph{Phys. Rev. Lett.} \textbf{77}, 2483--2486.


\bibitem[Frank \emph{et al.}, 2010]{fra10} \textsc{Frank}, A., \textsc{Bla\v{z}evi\'{c}}, A., \textsc{Grande}, P.L.,
\textsc{Harres}, K., \textsc{He{\ss}ling}, Th., \textsc{Hoffmann}, D.H.H., \textsc{Knobloch-Maas}, R., \textsc{Kuznetsov}, P.G.,
\textsc{N\"{u}rnberg}, F., \textsc{Pelka}, A., \textsc{Schaumann}, G., \textsc{Schiwietz}, G., \textsc{Sch\"{o}kel}, A.,
\textsc{Schollmeier}, M., \textsc{Schumacher}, D., \textsc{Sch\"{u}trumpf}, J., \textsc{Vatulin}, V.V., \textsc{Vinokurov},
O.A. \& \textsc{Roth}, M. (2010).
Energy loss of argon in a laser-generated carbon plasma.
\emph{Phys. Rev. E} \textbf{81}, 026401 (1--6).


\bibitem[Frank \emph{et al.}, 2013]{fra13} \textsc{Frank}, A., \textsc{Bla\v{z}evi\'{c}}, A., \textsc{Bagnoud}, V.,
\textsc{Basko}, M.M., \textsc{B\"{o}rner}, M., \textsc{Cayzak}, W., \textsc{Kraus}, D., \textsc{He{\ss}ling}, Th.,
\textsc{Hoffmann}, D.H.H., \textsc{Ortner}, A., \textsc{Otten}, A., \textsc{Pelka}, A., \textsc{Pepler}, D.,
\textsc{Schumacher}, D., \textsc{Tauschwitz}, An. \& \textsc{Roth}, M. (2013).
Energy loss and charge transfer of argon in a laser-generated carbon plasma.
\emph{Phys. Rev. Lett.} \textbf{110}, 115001 (1--5).


\bibitem[Gradshteyn \& Rizhik, 1980]{gra80} \textsc{Gradshteyn}, I.S. \& \textsc{Rizhik}, I.M. (1980).
\emph{Table of Integrals, Series and Products.} New York: Academic.


\bibitem[Hoffmann, 2008]{hof08} \textsc{Hoffmann}, D.H.H. (2008).
Intense laser and particle beams interaction physics toward inertial fusion.
\emph{Laser Part. Beams} \textbf{26}, 295--296.


\bibitem[Hoffmann \emph{et al.}, 2010]{hof10} \textsc{Hoffmann}, D.H.H., \textsc{Tahir}, N.A., \textsc{Udrea}, S.,
\textsc{Rosmej}, O., \textsc{Meister}, C.V., \textsc{Varentsov}, D., \textsc{Roth}, M., \textsc{Schaumann}, G.,
\textsc{Frank}, A., \textsc{Bla\v{z}evi\'{c}}, A., \textsc{Ling}, J., \textsc{Hug}, A., \textsc{Menzel}, J.,
\textsc{Hessling}, Th., \textsc{Harres}, K., \textsc{G\"{u}nther}, M., \textsc{El-Moussati}, S., \textsc{Schumacher}, D.
\& \textsc{Imran}, M. (2010).
High energy density physics with heavy ion beams and related interaction phenomena.
\emph{Contrib. Plasma Phys.} \textbf{50}, 7--15.


\bibitem[Hu \emph{et al.}, 2011]{hu11} \textsc{Hu}, Z.-H., \textsc{Song},
Y.-H., \textsc{Mi\v{s}kovi\'{c}}, Z.L. \& \textsc{Wang}, Y.-N. (2011).
Energy dissipation of ion beam in two-component plasma in the presence of
laser irradiation. \emph{Laser Part. Beams} \textbf{29}, 299--304.


\bibitem[Nellis, 2006]{nel06} \textsc{Nellis}, W.J. (2006).
Dynamic compression of materials: metallization of fluid hydrogen at high pressures.
\emph{Rep. Prog. Phys.} \textbf{69}, 1479--1580.


\bibitem[Nersisyan \& Akopyan, 1999]{ner99} \textsc{Nersisyan}, H.B. \& \textsc{Akopyan}, E.A. (1999).
Stopping and acceleration effect of protons in a plasma in the presence of an intense radiation field.
\emph{Phys. Lett. A} \textbf{258}, 323--328.


\bibitem[Nersisyan \& Deutsch, 2011]{ner11} \textsc{Nersisyan}, H.B. \& \textsc{Deutsch}, C. (2011).
Stopping of ions in a plasma irradiated by an intense laser field.
\emph{Laser Part. Beams} \textbf{29}, 389--397.


\bibitem[Nersisyan \& Deutsch, 2012]{ner12} \textsc{Nersisyan}, H.B. \& \textsc{Deutsch}, C. (2012).
Instabilities for a relativistic electron beam interacting with a laser-irradiated plasma.
\emph{Phys. Rev. E} \textbf{85}, 056414 (1--19).


\bibitem[Nersisyan \emph{et al.}, 2007]{ner07} \textsc{Nersisyan}, H.B., \textsc{Toepffer}, C.
\& \textsc{Zwicknagel}, G. (2007).
\emph{Interactions Between Charged Particles in a Magnetic Field: A Theoretical Approach to Ion
Stopping in Magnetized Plasmas.} Heidelberg: Springer.


\bibitem[Oguri \emph{et al.}, 2000]{ogu00} \textsc{Oguri}, Y., \textsc{Tsubuku}, K., \textsc{Sakumi}, A.,
\textsc{Shibata}, K., \textsc{Sato}, R., \textsc{Nishigori}, K., \textsc{Hasegawa}, J. \& \textsc{Ogawa}, M. (2000).
Heavy ion stripping by a highly-ionized laser plasma.
\emph{Nucl. Instrum. Methods Phys. Res. B} \textbf{161-163}, 155--158.


\bibitem[Piran, 2005]{pir05} \textsc{Piran}, T. (2005).
The physics of gamma-ray bursts.
\emph{Rev. Mod. Phys.} \textbf{76}, 1143--1210.


\bibitem[Roth \emph{et al.}, 2001]{rot01} \textsc{Roth}, M., \textsc{Cowan}, T.E., \textsc{Key}, M.H.,
\textsc{Hatchett}, S.P., \textsc{Brown}, C., \textsc{Fountain}, W., \textsc{Johnson}, J., \textsc{Pennington},
D.M., \textsc{Snavely}, R.A., \textsc{Wilks}, S.C., \textsc{Yasuike}, K., \textsc{Ruhl}, H., \textsc{Pegoraro},
F., \textsc{Bulanov}, S.V., \textsc{Campbell}, E.M., \textsc{Perry}, M.D. \& \textsc{Powell}, H. (2001).
Fast ignition by intense laser--accelerated proton beams.
\emph{Phys. Rev. Lett.} \textbf{86}, 436--439.


\bibitem[Roth \emph{et al.}, 2000]{rot00} \textsc{Roth}, M., \textsc{St\"{o}ckll}, C., \textsc{S\"{u}}{\ss}, W.,
\textsc{Iwase}, O., \textsc{Gericke}, D.O., \textsc{Bock}, R., \textsc{Hoffmann}, D.H.H., \textsc{Geissel}, M.
\& \textsc{Seelig}, W. (2000).
Energy loss of heavy ions in laser-produced plasmas.
\emph{Europhys. Lett.} \textbf{50}, 28--34.


\bibitem[St\"{o}ckl \emph{et al.}, 1996]{sto96} \textsc{St\"{o}ckl}, C., \textsc{Frankenheim}, O.B., \textsc{Roth}, M.,
\textsc{Su}{\ss }, W., \textsc{Wetzler}, H., \textsc{Seelig}, W., \textsc{Kulish}, M., \textsc{Dornik}, M., \textsc{Laux}, W.,
\textsc{Spiller}, P., \textsc{Stetter}, M., \textsc{St\"{o}we}, S., \textsc{Jacoby}, J. \& \textsc{Hoffmann}, D.H.H. (1996).
Interaction of heavy ion beams with dense plasmas.
\emph{Laser Part. Beams} \textbf{14}, 561--574.


\bibitem[Tabak \emph{et al.}, 1994]{tab94} \textsc{Tabak}, M.,
\textsc{Hammer}, J., \textsc{Glinsky}, M.E.,
\textsc{Kruer}, W.L., \textsc{Wilks}, S.C., \textsc{Woodworth}, J.,
\textsc{Campbell}, E.M., \textsc{Perry}, M.D. \& \textsc{Mason}, R.J. (1994).
Ignition and high gain with ultrapowerful lasers.
\emph{Phys. Plasmas} \textbf{1}, 1626--1634.


\bibitem[Tahir \emph{et al.}, 2005]{tah05} \textsc{Tahir}, N.A., \textsc{Deutsch}, C., \textsc{Fortov}, V.E.,
\textsc{Gryaznov}, V., \textsc{Hoffmann}, D.H.H., \textsc{Kulish}, M., \textsc{Lomonosov}, I.V., \textsc{Mintsev}, V.,
\textsc{Ni}, P., \textsc{Nikolaev}, D., \textsc{Piriz}, A.R., \textsc{Shilkin}, N., \textsc{Spiller}, P., \textsc{Shutov}, A.,
\textsc{Temporal}, M., \textsc{Ternovoi}, V., \textsc{Udrea}, S. \& \textsc{Varentsov}, D. (2005).
Proposal for the study of thermophysical properties of high-energy-density matter using current and future
heavy-ion accelerator facilities at GSI Darmstadt.
\emph{Phys. Rev. Lett.} \textbf{95}, 035001 (1--4).


\bibitem[Tavdgiridze \& Tsintsadze, 1970]{tav70} \textsc{Tavdgiridze}, T.L. \& \textsc{Tsintsadze}, N.L. (1970).
Energy losses by a charged particle in an isotropic plasma located in an external high frequency electric field.
\emph{Zh. Eksp. Teor. Fiz.} \textbf{58}, 975--978 [English translation: \emph{Sov. Phys. JETP} \textbf{31}, 524--525 (1970)].


\bibitem[Wang \emph{et al.}, 2002]{wan02}
\textsc{Wang}, G.-Q., \textsc{Song}, Y.-H., \textsc{Wang}, Y.-N. \& \textsc{Mi\v{s}kovi\'{c}}, Z.L. (2002).
Influence of a laser field on Coulomb explosions and stopping power for swift molecular ions
interacting with solids.
\emph{Phys. Rev. A} \textbf{66}, 042901 (1--11).


\bibitem[Wang \emph{et al.}, 2012]{wan12}
\textsc{Wang}, G.-Q., \textsc{E}, P., \textsc{Wang}, Y.-N., \textsc{Hu}, Z.-H., \textsc{Gao}, H.,
\textsc{Wang}, Y.-C., \textsc{Yao}, L., \textsc{Zhong}, H.-Y., \textsc{Cheng}, L.-H., \textsc{Yang}, K.,
\textsc{Liu}, W. \& \textsc{Xu}, D.-G. (2012).
Influence of a strong laser field on Coulomb explosion and stopping power
of energetic H$^{+}_{3}$ clusters in plasmas.
\emph{Phys. Plasmas} \textbf{19}, 093116 (1--5).


\bibitem[Zwicknagel \emph{et al.}, 1999]{zwi99} \textsc{Zwicknagel}, G., \textsc{Toepffer}, C.
\& \textsc{Reinhard}, P.-G. (1999).
Stopping of heavy ions in plasmas at strong coupling.
\emph{Phys. Rep.} \textbf{309}, 117--208.

\end{thebibliography}
\end{document}